\documentclass[twocolumn,
aps,
prb,
showpacs,
amsmath,
amssymb,
floatfix,
raggedbottom,
nobalancelastpage,
reprint,
superscriptaddress,
noeprint]{revtex4-1}

\pdfoutput=1

\usepackage{xcolor,mathtools,bm,dsfont}
\definecolor{darkblue}{RGB}{0,0,150}
\definecolor{nightblue}{RGB}{0,0,100}

\usepackage{graphicx}
\usepackage[
colorlinks,
citecolor=darkblue,
linkcolor=darkblue,
urlcolor=nightblue]{hyperref}

\usepackage[english]{babel}
\usepackage[babel,kerning=true,spacing=true]{microtype}

\newcommand{\dd}{\mathrm{d}}

\newcommand{\inv}{\ensuremath{^{-1}}}
\DeclareMathOperator{\sgn}{sgn}

\bibpunct{[}{]}{,}{n}{}{}
\makeatletter
\def\NAT@def@citea{\def\@citea{\NAT@separator}}
\makeatother

\begin{document}

\title{
Ballistic and hydrodynamic magnetotransport in narrow channels
}

\author{Tobias Holder}
\email{tobias.holder@weizmann.ac.il}
\affiliation{Department of Condensed Matter Physics,
Weizmann Institute of Science,
Rehovot 7610001, Israel}
\author{Raquel Queiroz}
\email{raquel.queiroz@weizmann.ac.il}
\affiliation{Department of Condensed Matter Physics,
Weizmann Institute of Science,
Rehovot 7610001, Israel}
\author{Thomas Scaffidi}
\affiliation{Department of Physics, University of California, Berkeley, California 94720, USA}
\author{Navot Silberstein}
\affiliation{Department of Condensed Matter Physics,
Weizmann Institute of Science,
Rehovot 7610001, Israel}
\author{Asaf Rozen}
\affiliation{Department of Condensed Matter Physics,
Weizmann Institute of Science,
Rehovot 7610001, Israel}
\author{Joseph A.\ Sulpizio}
\affiliation{Department of Condensed Matter Physics,
Weizmann Institute of Science,
Rehovot 7610001, Israel}
\author{Lior Ella}
\affiliation{Department of Condensed Matter Physics,
Weizmann Institute of Science,
Rehovot 7610001, Israel}
\author{Shahal Ilani}
\affiliation{Department of Condensed Matter Physics,
Weizmann Institute of Science,
Rehovot 7610001, Israel}
\author{Ady Stern}
\affiliation{Department of Condensed Matter Physics,
Weizmann Institute of Science,
Rehovot 7610001, Israel}

\date{\today}

\begin{abstract}
An increasing number of low carrier density materials exhibit a surprisingly large transport mean free path due to inefficient momentum relaxation. Consequently, charge transport in these systems is markedly non-ohmic but rather ballistic or hydrodynamic, features which can be explored by driving current through narrow channels.
Using a kinetic equation approach we theoretically investigate how a non-quantizing magnetic field discerns ballistic and hydrodynamic transport, in particular in the spatial dependence of the transverse electric field, $E_y$: We find that $E_y$ is locally enhanced when the flow exhibits a sharp directional anisotropy in the non-equilibrium density. 
As a consequence, at weak magnetic fields, the curvature of $E_y$ has opposite signs in the ballistic and hydrodynamic regimes. 
Moreover, we find a robust signature of the onset of non-local correlations in the form of distinctive peaks of the transverse field, which are accessible by local measurements.
Our results demonstrate that a purely hydrodynamic approach is insufficient in the Gurzhi regime once a magnetic field is introduced.
\end{abstract}

\maketitle

\section{Introduction} 
From high-mobility heterostructures based on GaAs~\cite{deJong1995} to the advent of materials with low carrier density like Graphene~\cite{Mueller2008,Fritz2008,Mueller2009,Bistritzer2009,Mendoza2011,Mendoza2013,Narozhny2015,Briskot2015,Lucas2016} and later Weyl- and Dirac semimetals~\cite{Armitage2018}, formerly elusive regimes of hydrodynamic and ballistic flow are now at the reach of present day experiments~\cite{Gurzhi1968,Beenakker1991,Bandurin2016,KrishnaKumar2017,Moll2016,Gooth2017,Berdyugin2019}.
The activity surrounding transport in clean systems has also reinvigorated the study of viscous, i.~e. non-local properties of collective electronic motion, but until a few years ago, the real-space properties of hydrodynamically flowing electrons in a magnetic field was all but conjectured. This is due to the difficulties associated with a local measurement of electronic flow, which is sensitive to the placement of contacts and gates in traditional experimental approaches. 
As a consequence, considerable effort has gone into the diagnostics of non-ohmic electron flow and parity violating hydrodynamics in general~\cite{Andreev2011,Tomadin2014,Torre2015,Levchenko2017,Svintsov2018,Ho2018,Lucas2014,Sherafati2016,Ganeshan2017,Guerrero-Becerra2018,Gusev2018}. More recently, the same phenomenology was discussed for some active classical fluids~\cite{Lapa2014,Tuegel2017,Banerjee2017,Soni2018,Souslov2019}. Nevertheless, a clear characteristic of the flow that would signal the onset of non-local correlations ranging beyond those associated with hydrodynamic flow has remained elusive. Consequently, it has so far been unclear how non-local transport can be diagnosed reliably in the intermediate regime where all length scales are of comparable size.

\begin{figure}
    \centering
    \includegraphics[width=\columnwidth]{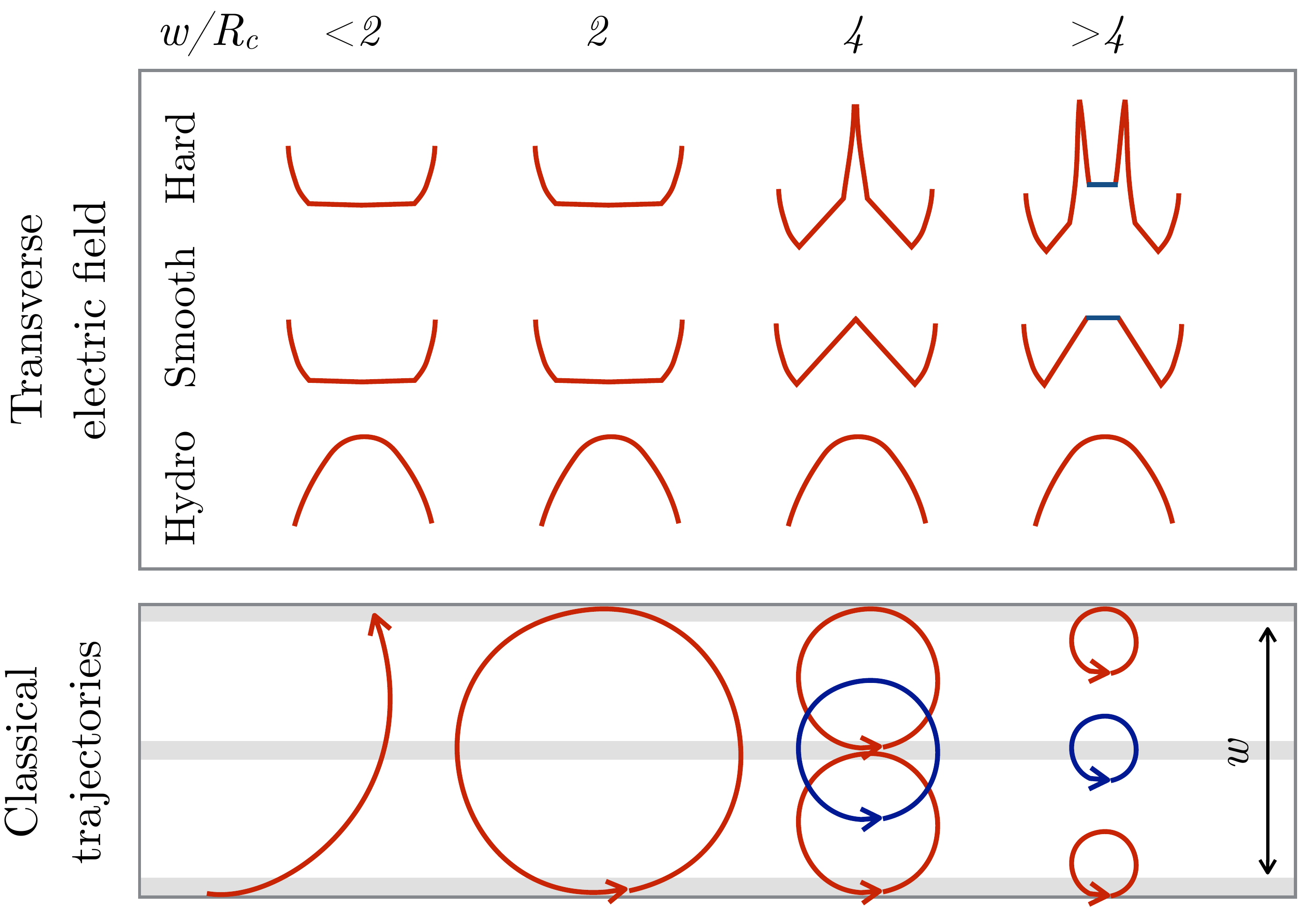}
    \caption{Top: Schematic illustrations of the spatial profiles of the transverse electric field $E_y$ at different ratios $w/R_c$. The first row shows the ballistic regime ($\ell_{ee}>\ell_0>w$), and the second one the hydrodynamic case ($\ell_{0}>w>\ell_{ee}$). The boundary divergence leads to a positive curvature for weak magnetic fields; In the hydrodynamic regime the flow is expected to be parabolic independently of magnetic field. Bottom: Classical trajectories relevant for the ballistic regime: Bulk trajectories (blue) determine the flow in the middle of the channel for $w/R_c>4$, while edge trajectories (red) dominate the flow for weaker fields. In the ballistic regime the divergent contribution to $E_y$ is a result of boundary tangent trajectories which spend a disproportionally long time at the boundary and at $2R_c$.}
    \label{fig:scheme}
\end{figure}

To observe viscous effects in transport, the most often considered geometry is a narrow channel with width $w$, in which the electrons flow in a fashion comparable to classical fluids in a pipe. Until very recently, such a Poiseuille-like current profile has been unequivocally associated with the presence of hydrodynamic flow.
However, it has now become possible to precisely determine the spatial profile of the electrostatic environment of flowing electrons thanks to new experimental techniques~\cite{Ella2018,Sulpizio2019}.
These results suggest that a spatial modulation of the electron drift velocity can be explained by several non-local effects other than hydrodynamic viscous forces. In particular, a clean material with ballistic flow -- nearly non-interacting with a long electron mean free path -- also exhibits highly non-local features resembling those of viscous flow. 
It is thus necessary to not only understand the role of viscous effects in clean electron fluids but also to resolve the ballistic regime and the crossover between the two. 
We point out that the current density will invariably decrease near a diffusive boundary, merely as a result of momentum dissipation, making this quantity rather unsuitable for a precise differentiation of the various flow regimes. Therefore, in the following we do not try to resolve small differences in the current density profile. Instead, we propose a readily measurable local quantity which is very sensitive to the type of non-local correlation present in the flow: The transverse electric field.

To this end, we solve the kinetic equation in the presence of a confining channel, impurity scattering, electron-electron scattering and a non-quantizing, out-of-plane, magnetic field. 
This allows us to improve on the results obtained in a hydrodynamic paradigm where a solution is given by an angular moment expansion~\cite{Alekseev2016,Falkovich2017,Delacretaz2017}; in particular we construct an analytical solution in the little studied ultraballistic limit at infinite disorder ($\ell_0$) and electron-electron interaction ($\ell_{ee}$) mean free paths. In the latter, local quantities like current density and electric field can be connected to classical electron trajectories in cyclotron motion, as schematized in Fig.~\ref{fig:scheme}. Remarkably, we find in this limit a divergent contribution to the transverse electric field around the edges of the channel, which induces an overall positive curvature in its profile. This upturn of the electric field is present down to $\ell_0\sim w$. 
As the reason we identify a drastically enhanced Hall response in the ballistic case due to the large stress, i.e. the large velocity anisotropy that accompanies cyclotron orbits which tangentially touch a diffusive boundary. These orbits spend a disproportionately long time in this region, in comparison to other trajectories. 
Since this strong signal does not appear in viscous flow, we propose to use the profile of the Hall electric instead of the current density to probe inhomogeneous electron flow.

In particular, tuning the applied magnetic field to the special value of $w=4R_c$, where $R_c$ is the radius of the cyclotron orbit, skipping orbits occupy exactly one half of the channel, with the largest orbits reaching up to the center of the channel. 
In this case, the electron flow at the middle of the channel is influenced particularly strongly by the boundary and can serve as a sensitive indicator for the type of non-locality (hydrodynamic vs. ballistic) present in the sample. The expected profiles along the channel width are schematically depicted in Fig~\ref{fig:scheme}. 
We suggest to use this peak in the transverse electric field for $w=4R_c$ as an accessible marker for the breakdown of viscous flow.

Unlike at zero magnetic field, this ballistic-hydrodynamic crossover is not captured by the hydrodynamic formalism, which only accounts for viscous forces in the form of velocity gradients. This applies in particular to the so called Gurzhi regime where the flow is viscous ($\ell_{ee}<w$) but the bulk momentum relaxing mean free path exceeds the sample dimensions ($\ell_0>w$) such that their geometric mean $\sqrt{\ell_{ee}\ell_0}$ is comparable to $w$. Contrary to conventional wisdom, an inhomogeneous current profile in this regime is not an unambiguous signature of viscous forces.

The remaining sections of the paper are organized as follows. In Sec.~\ref{sec:two} we present our results found for both hydrodynamic and ballistic flow; particular attention is hereby payed to the conceptual assumptions that enter the modeling of the transverse electric field profile and the longitudinal current profile. 
The calculation is documented in two steps. In Sec.~\ref{sec:three}, the general kinetic equation approach is explained, together with the features it shares with the hydrodynamic formalism and the points at which it deviates from the latter. 
The general solution to the Boltzmann equation for viscous and ballistic flow is explained in Sec.~\ref{sec:four}, proceeding constructively mainly by studying the distribution function itself, and only then analyzing the actual observables like transverse electric field and current density based on the insights from the complete information about the non-equilibrium properties as provided by the distribution function. 
We conclude in Sec.~\ref{sec:five}.

\section{Summary of results}
\label{sec:two}
We consider a homogeneous channel of width $w$ in which an electron liquid with distribution function $f(\bm{r},\bm{p},t)$ is flowing.  The kinetic equation is
\begin{align}
\partial_{t} f
+\bm{v}\cdot\nabla_{\bm{r}} f
+e(\bm{E}+\bm{v}\times\bm{B})\cdot\nabla_{\bm{p}} f&=\mathcal{I}(f),
\label{eq:boltzmann}
\end{align}
where in the following we take the relaxation time approximation for the collision integral $\mathcal{I}(f)$.
At low temperatures, only processes close to the Fermi surface are important and it is sufficient to parametrize the nonequilibrium part by
$f(\bm{r},\bm{p},t)=f_0-E_F(\partial_{\epsilon} f_0) h(y,\theta),$ 
with $\theta$ being the angle along the Fermi surface with Fermi energy $E_F$. In terms of the smooth nonequilibrium distribution function $h(y,\theta)$ the kinetic equation becomes
\begin{align}
&R_c\inv\partial_\theta h(y,\theta)
+\sin\theta\partial_y h(y,\theta)
=-\frac{ h(y,\theta)}{\ell}+
F(y,\theta).
\label{eq:main1}
\end{align}
The inhomogeneous term $F(y,\theta)$ is conveniently written using even and odd angular momentum components of $h$ denoted $h^{e,o}_l(y)$ (where $l=0,1,2,\dots$) 
\footnote{The angular momentum components of the distribution function are defined by $h^{e}_l(y)=\pi^{-1}\int  h(\theta,y)\cos (l\theta) d\theta$ and $h^{o}_l(y)=\pi^{-1}\int  h(\theta,y)\sin (l\theta) d\theta$},
\begin{align}
F(y,\theta)
&=E_x\cos\theta-
E_y(y)\sin\theta
\notag\\
&+\frac{h_0(y)}{\ell}
+\frac{ h_1^e(y)\cos\theta+h_1^o(y)\sin\theta}{\ell_{ee}}
\label{eq:main2},
\end{align}
the last two terms ensure conservation of particle density and conservation of momentum under electron-electron interactions~\cite{deJong1995,Landaubook9}. 
Using this scattering integral, the kinetic approach contains both the fast, local equilibration in non-conserved degrees of freedom and the slow, diffusive equilibration of conserved quantities, such that it can successfully describe the multiple crossovers between ohmic, hydrodynamic and ballistic flow. In particular, it describes how electron-electron interaction conserves the momentum of the center of mass, while distributing this momentum between all electrons. Note that we use units such that $E_x$, $E_y$ are wavenumbers and chose signs making $E_x,E_y,R_c>0$. By multiplying Eq.~\eqref{eq:main1} by $\sin{\theta}$ and integrating over $\theta$, one can immediately check that the transverse electric field is self-consistently given by
\begin{align}
    E_y(y)=\tfrac{1}{R_c}h_1^e(y)+\tfrac{1}{2}\partial_y h_2^e(y)-\partial_yh_0(y).\label{eq:selfconsistentEy}
\end{align} 
A solution to Eqs.~(\ref{eq:main1},\ref{eq:main2}) can be found by employing the method of characteristics by rewriting the kinetic equation as an evolution equation to the flow characteristics, which correspond to the classical particle trajectories when the effective mean free path $\ell=\ell_0\ell_{ee}/(\ell_0+\ell_{ee})$ is infinite. 
In the kinetic approach, the boundary scattering is implemented by rescaling the value of the non-equilibrium distribution function by a factor $0\leq r_h\leq 1$ upon hitting the boundary. Here, $r_h=0$ corresponds to diffusive scattering which completely randomizes momenta, while $r_h=1$ is perfectly specular. 

\subsection{Hydrodynamic regime}
The distribution function obtained from solving the kinetic equation is generally not a smooth function, showing discontinuities at the boundary as well as between regions dominated by bulk or boundary trajectories. It is a smooth function in the hydrodynamic regime, where momentum-conserving electron-electron interactions result in a small quasiparticle mean free path $\ell_{ee}\ll w$. In this case the distribution function is well approximated by a truncation in angular momentum components $h^{e,o}_l(y)$. 
We note that the precise mechanism which leads to $\ell_{ee}\ll w$ is not important in the present context, except for it to be an intrinsic (bulk) mechanism, like electron-electron scattering or momentum conserving phonon scattering~\footnote{The most famous example for a Fermi liquid with strong interactions is undoubtedly Graphene at finite density, where both theory and experiment estimate $\ell_{ee}\sim 1\mathrm{\mu m}\ll w$ at temperature $T=75\mathrm{K}$~\cite{Schuett2011,Svintsov2018,Ho2018,Bandurin2016,KrishnaKumar2017}.}. An approximation up to second order in angular moments ($l=2$) is identical to the previously used hydrodynamic treatment, while a truncation at $l=3$ allows to impose boundary conditions both on the first and second angular modes of the distribution function, related to current and stress~\footnote{The stress tensor is closely related to $h_2^{e/o}$. Disregarding charging effects, stress tensor and second moments can be used interchangeably.}. 
Here, we show the difference between both truncations and derive the generalized boundary conditions on the stress components in terms of the longitudinal and transverse resistances. The solution of the second order equations for the transverse electric field, valid deep in the hydrodynamic regime has been calculated before~\cite{Alekseev2016}
\begin{align}
    E_y=
    \frac{E_x}{R_c}
    \left(\ell_0-\left(2 \ell+\ell_0\right) (1-r_j) 
    \frac{\cosh (2y/\ell_c)}
    {\cosh(w/\ell_c)}
    \right), \label{eq:result2harm}
\end{align}
where $R_c$ is the cyclotron radius and $\ell_c$ is the effective mean free path in the presence of a magnetic field with $\ell_c^2=\ell_0\ell R_c^2/(4\ell^2+R_c^2)$~\footnote{We point out that at zero magnetic field this length approaches the Gurzhi length, $\ell_c\rightarrow\sqrt{\ell\ell_0}$.}. The  ratio $r_j=h_1^e(w/2)/h_{1,bulk}^e$ of current density at the boundary divided by the bulk current density (namely, the current density in the limit of an infinitely wide channel) is assumed in this formulation to be a material parameter. It can take values $0\leq r_j \leq 1$ , with $r_j=1$ corresponding to a specular reflection while $r_j=0$ refers to a no-slip boundary condition~\cite{Kiselev2018}. In the literature, $r_j$ is usually assumed freely adjustable. 
This is questionable in the presence of weak magnetic fields. By comparison with the numerical solution of the kinetic equation, we find that the moment expansion can successfully describe hydrodynamic flow as long as $\ell_0<w$, but using no-slip boundary conditions has several unexpected consequences once $\ell_0>w$. 
For example, we observe in Eq.~\eqref{eq:result2harm} that the transverse electric field changes its sign in a small layer close to the wall if
$r_j<{2\ell_{ee}}/({\ell_{0}+3\ell_{ee}})$.
Thus even deep in the hydrodynamic regime ($\ell_{ee}\to 0$), using a no-slip boundary conditions will produce such a sign reversal.
Furthermore, when $\sqrt{\ell\ell_0}\gg w$ and also $\ell>w/\sqrt{6}$, in a weak magnetic field this sign change carries over to the 
Hall voltage. We reiterate that the sign reversal is not reproduced in the kinetic approach for any choice of $r_h$. 
The issue can be understood by looking at the extreme case $\ell_{ee},R_c\to\infty$ where the kinetic equation has an exact solution. Taking fully diffusive walls ($r_h\!=\!0$), we find that for values of $\ell_0\lesssim w$, the current density at the boundary is approximately half of the bulk current density, so a no slip boundary condition is far too restrictive. For larger $l_0$ or at finite $\ell_{ee}$, this ratio, while smaller, remains nonzero and the associated lower bound on $r_j>0$ needs to be inferred from the more complete kinetic approach.
In short, the hydrodynamic approach, and Eq.~\eqref{eq:result2harm} are only reliable deep in the hydrodynamic regime and might lead to wrong results once $\ell\sim w$.

\subsection{Ballistic regime in weak magnetic field}  

\begin{figure}
    \centering
    \includegraphics[width=\columnwidth]{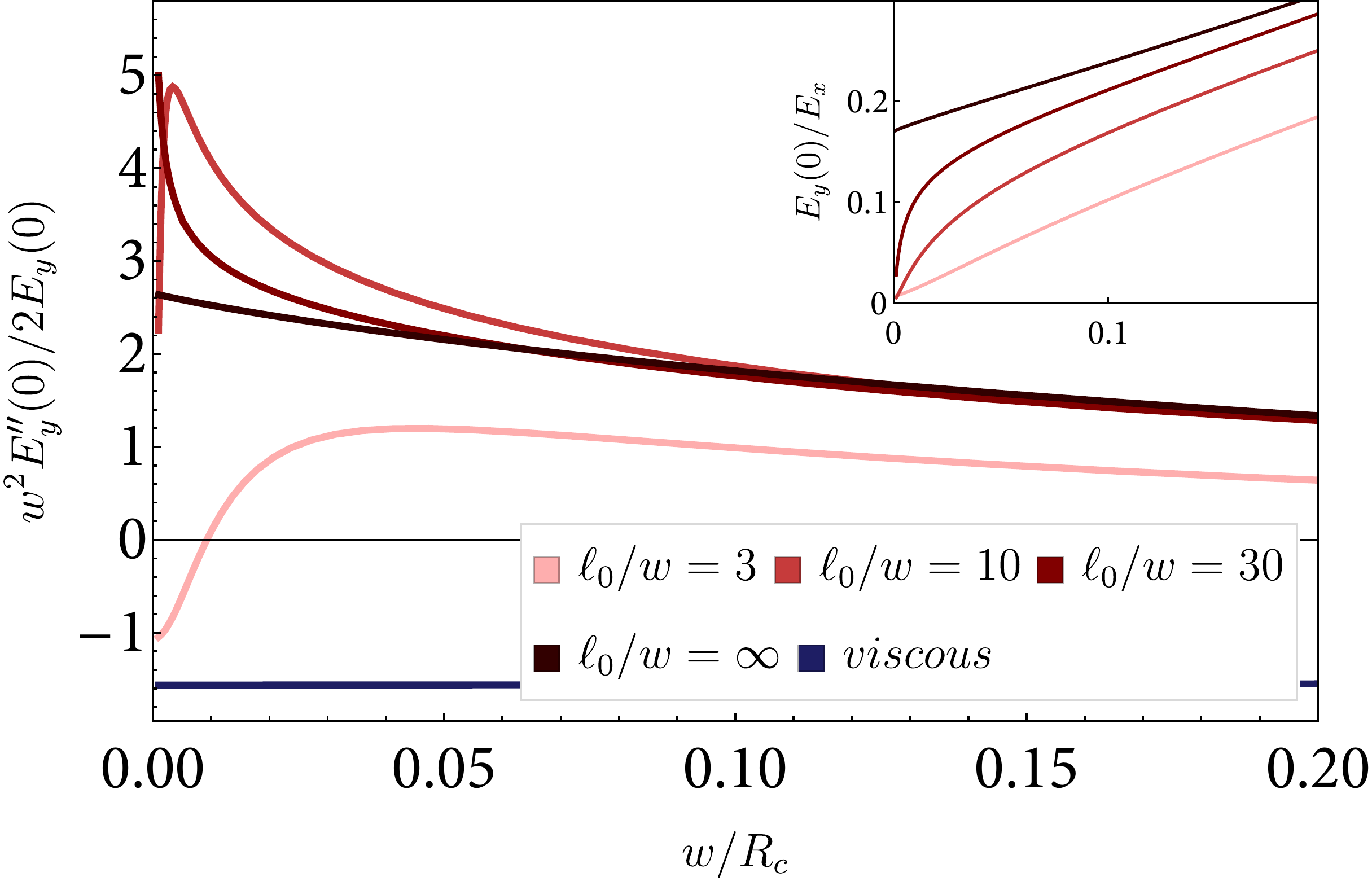}
    \caption{Normalized curvature of the transverse electric field in the center of the channel if $\ell_{ee}\rightarrow\infty$. For $l_0>R_c$ and unless the disorder mean free path becomes comparable to the channel width, $l_0\sim w$, the inverted (positive) sign of the curvature is robust. For comparison, a typical curve in the hydrodynamic regime is also plotted using $\ell_{ee}=0.3w$ (blue), which shows that the curvature is negative and almost independent of $R_c$. Inset: Saturation of the Hall electric field at a non-zero value in the ultra-ballistic limit.}
    \label{fig:smallBEycurv}
\end{figure}

The viscous and ballistic resistances differ fundamentally when subjected to an out-of-plane magnetic field~\cite{Scaffidi2017,Alekseev2018}. However, it remains challenging to extract unique signatures of hydrodynamic transport from such data. We find that particularly telling in this respect is the spatial profile of the transverse electric field $E_y$. At weak fields ($R_c>w$), for ballistic transport the profile is nearly parabolic with positive curvature, while it retains a negative curvature for hydrodynamic flow (cf. Fig.~\ref{fig:smallBEycurv}).

This is related to the different sources of stress, i.e. differences in the origin of the angular components with $l=2$. In viscous flow, the derivative $\partial_yh_2^e$ which enters $E_y$ in Eq.~\eqref{eq:selfconsistentEy} represents a small correction which is proportional to the curvature of the velocity profile. In contrast, in ballistic transport $E_y$ receives a large contribution from $\partial_yh_2^e$ which is now more appropriately interpreted as an imbalance between the density of orbits which just about touch the boundary compared to skipping orbits which are surppressed by the diffusive boundary scattering (cf. Fig.~\ref{fig:scheme}).
As long as $2R_c>w$ the high stress only applies near the boundary, where orbits which graze the boundary retain a sizable number density while perpendicularly hitting trajectories lose momentum. This effect completely overpowers the normal Drude term in $E_y$ and leads to the inverted curvature in the transverse electric field. Note that this mechanism is neither possible in the absence of a magnetic field, as proximate parallel trajectories also hit the boundary in close proximity, nor can it happen in hydrodynamic flow, where e-e interactions constantly equilibrate neighbouring orbits. 
To reiterate, even if two fluids have exactly the same current profile, the Hall electric field can be dissimilar due to the presence of higher angular moments in the distribution function. In essence, while all moments of the distribution function are coupled through the kinetic equation, currents ($\ell=1$) and viscous terms ($\ell=2$) are affected differently by the higher order angular momenta with only the latter being significantly modified by the increasingly non-local correlations in the ballistic regime.

These observations can be made precise by directly investigating the case $w<R_c<\ell_0<\ell_{ee}$. Using $r_h\!=\!0$, the distribution function can be divided into two symmetric smooth parts. The building block is ${\tilde h}(y,\theta)$  defined for angles between $\theta_i(y)=\arccos(1-(w/2+y)/R_c)$ and $\theta_f(y)=2\pi-\arccos((w/2-y)/R_c-1)$, which indicate the trajectories that originate in the lower boundary. Adding also the trajectories that originate from the upper boundary yields $h(y,\theta)={\tilde h}(+y,\theta)+{\tilde h}(-y,\pi-\theta)$.
Expanding this expression for $l_0\to\infty$~(cf. Sec.~\ref{sec:A}), the result is
\begin{align}
    {\tilde h}(y,\theta)&=
    E_x R_c\left(
    \sin\theta-\sqrt{1-\left(\frac{\tfrac{w}{2}+y}{R_c}+\cos\theta\right)^2}
    \right).
    \label{eq:infinitel0}
\end{align}
Here, we already disregarded the feedback of the transverse electric field in the kinetic equation, which is non-singular. The large contribution to  stress at the boundary is completely sourced from the square root in Eq.~\eqref{eq:infinitel0}, which leads to a logarithmic divergence at the boundaries upon angular integration. 
Integrating the distribution function to obtain the angular moments, it turns out that additionally the Hall voltage retains a non-zero value for vanishing magnetic field (Fig.~\ref{fig:smallBEycurv}). The enhanced Hall response is due to trajectories which travel large distances on a slightly bent path along the channel and thus it emerges for $w<R_c<\ell_0$. 
This behavior in the extreme ballistic limit is known under the name ``last plateau''~\cite{Beenakker1989}.
For the normalized curvature in this regime we find $E_y''(0)/2E_y(0)\approx 2.65 / w^2>0$.
However, the last plateau breaks down once $R_c>\ell_0$, and the usual zero magnetic field limit is recovered beyond this point (Fig.~\ref{fig:smallBEycurv}). 

\subsection{Critical field $w/R_c=4$}
We now turn to the critical value of magnetic field $w/R_c=4$. We find this value to offer the clearest insight into magnetotransport and the properties of the boundary: it corresponds to cyclotron orbits which completely decouple the influence of the two channel walls. In this case, the middle of the channel shows a remarkable sensitivity to nonlocal transport signatures, which vanish in the hydrodynamic limit.

\begin{figure}
    \centering
    \includegraphics[width=\columnwidth]{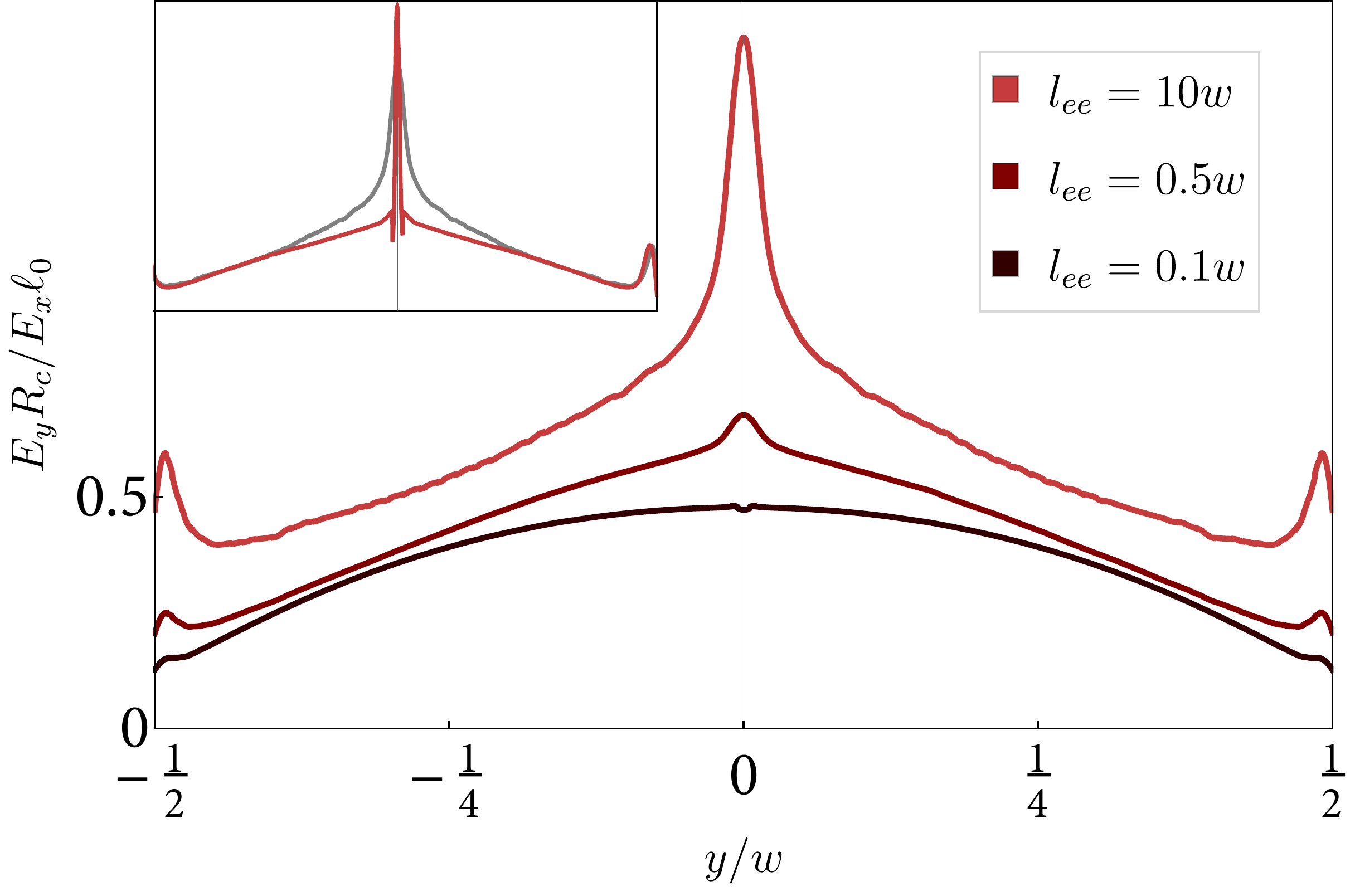}
    \caption{Transverse electric field for different interaction mean free paths $\ell_{ee}$. The profile becomes increasingly parabolic with increasing interaction strength, indicating the crossover from ballistic to hydrodynamic transport. The Gurzhi length is equal to the channel width for $\ell_{ee}=0.21w$. In this regime, at $\ell_{ee}=0.5w$, remnants of ballistic transport are still visible. Here, the effective mean free path is $\ell>6^{-1/2}w$, which implies that the mode expansion suited for the hydrodynamic limit is not applicable. The remaining parameters are $l_0=5w$, $w/R_c=4$. In the inset the solution for diffusive bounces (gray) is compared with a narrow absorptive boundary layer (red), the latter suppresses the central peak, leaving flat shoulders. A cusp in the center of the channel remains.}
    \label{fig:crosstohydro}
\end{figure}

We calculate the current profile using a boundary model with fully diffusive walls and find a distinctive dip of the current density at the center of the channel. The profile of $E_y$ is even more sensitive as it contains the term $\partial_y h^e_2$ related to stress. As explained before, the termination points of orbits which merely touch the boundary show peaks in $E_y$. At $w/R_c=4$ these extremal cyclotron orbits from both boundaries reach precisely up to the middle of the channel, producing a very prominent peak there (Fig.~\ref{fig:crosstohydro}). This effect is present as long as the orbits remain relatively unperturbed, that is as long as $\sqrt{\ell\ell_0}>w$, therefore the disappearance of the peaks in $E_y$ mark the transition point to proper hydrodynamic transport.  
As we showed, the geometric mean $\sqrt{\ell\ell_0}\sim w$ actually exhibits characteristics of ballistic transport. What is referred to as Poiseuille flow of electrons in this regime is therefore more appropriately identified as a ballistic parabolic profile. Hydrodynamic flow is only reached if not only $l_0\gg w$ but also  $\sqrt{\ell\ell_0}\ll w$.

Since the most prominent feature of the ballistic-hydrodynamic crossover is the divergent contribution of stress from orbits who just touch the boundary, one has to be careful about the robustness of such features. Thus, we explicitly test how the results change when the current is completely annihilated at the boundary (\emph{superdiffusive} boundary condition). While neither theory nor experiment indicate that such a drastic boundary friction is present in the intermediate regimes of transport~\cite{Kiselev2018}, it presents a useful extremal case where the discontinuities in the distribution function are most effectively suppressed. Yet, we find that the distinctive difference in the profile of $E_y$ between hydrodynamic and ballistic transport survives, although with weaker magnitude (cf. inset in Fig.~\ref{fig:crosstohydro}).
We thus conclude that at the critical value $w/R_c=4$, the profile of $E_y$ exhibits a robust signal of ballistic transport at the center of the channel, which assumes the form of a cusp or even a narrow peak. This form is retained in the entire Gurzhi regime.

\section{Kinetic equation: Mode expansion}
\label{sec:three}
We already presented the profile of the transverse electric field, Eq.~\eqref{eq:result2harm} for hydrodynamic transport. In the following we briefly rederive this result with a focus on microscopic length scales, commenting on the limitations which appear at weak magnetic fields, and also generalizing the formalism to next-to-leading order in the derivative expansion.

From the theoretical side, the current density profile in the interaction dominated regime was worked out in~\cite{Alekseev2016,Falkovich2017,Delacretaz2017}.
In essence, the isochoric (i.e. divergence-free) flow is given by the solution to a Poisson-like equation 
\begin{align}
    \Delta\left(\Delta-\frac{1}{\gamma^2}\right)\psi&=0
    \label{eq:hydro1}
\end{align}
for the scalar stream function $\psi$, which in turns defines the flow field $\bm{v}$ by the relation  $v_i=\epsilon_{ij}\partial_j\psi$. The length $\gamma$ indicates the distance over which perturbations of the flow are alleviated.   
Applied to a channel geometry, the solution of this equation is straightforward, yielding a hyperbolic flow profile in terms of hydrodynamic properties. The same solution can be constructed based on a kinetic equation approach, where the system's properties are now encapsulated by various mean free paths. For later convenience, here we retrace this latter approach.

\subsection{Definitions}
Unless otherwise noted, the units are chosen such that $k_B=\hbar=1$.
Starting from the general form of the kinetic equation as stated in Eq~\eqref{eq:boltzmann},
we assume in the following a free fermion dispersion $\epsilon(\bm{k})=k^2/2m$, the number density is $n$.
At low temperatures, only processes close to the Fermi surface are important and it is useful to parametrize the nonequilibrium part with a smooth dimensionless function
\begin{align}
f(\bm{r},\bm{p},t)&=f_0-E_F(\partial_{\epsilon} f_0) h(\bm{r},\theta,t),
\end{align}
with $\theta$ being the angle along the Fermi surface.
Due to the channel geometry, derivatives with respect to $x$ in the channel direction vanish. Thus in the steady state only a dependence on the perpendicular coordinate $y$ and the angle $\theta$ remains.
The force term becomes in cylindrical coordinates
\begin{align}
\bm{F}\cdot\nabla_{\bm{p}} f&=e(\bm{E}+\bm{v}\times\bm{B})\cdot(\hat p\partial_p+p\inv\hat\theta\partial_\theta)f\notag\\&=E_F\nu(v\tilde E_x\cos\theta+v\tilde E_y\sin\theta+ \omega_c\partial_\theta h),
\end{align}
where we introduced the wavenumbers 
$\tilde E_i\!=\!e E_i/E_F$ and the cyclotron frequency $\omega_c\!=\!e B / E_F$, $\nu$ is the density of states. From now on we drop the tildes and substitute $E_x\to-E_x$. Then, the kinetic equation assumes the simplified form
\begin{align}
&v\sin\theta\partial_y h(y,\theta)
-vE_x\cos\theta
+vE_y\sin\theta
\notag\\&\quad
+\omega_c\partial_\theta h(y,\theta)=\mathcal{I}(h).
\label{eq:kinetic}
\end{align}

\subsection{Hydrodynamic limit}
The hydrodynamic solution is reproduced by expanding the nonequilibrium distribution function $h(y,\theta)$ in its angular momentum components $h_l(y)$. Such an approximation is valid as long as the distribution function remains smooth.
Up to second order the even (e) and odd (o) moments are
\begin{align}
 h(y,\theta)&=h_0(y)+ h^e_1(y)\cos\theta+ h^o_1(y)\sin\theta
 \notag\\&\quad
 + h^e_2(y)\cos2\theta+ h^o_2(y)\sin2\theta,
\end{align}
where $h^e_1$ and $h^o_1$ are proportional to the longitudinal and transverse currents, and $h^e_2$ and $h^o_2$ correspond to nematic deformations of the Fermi surface due to stress in the liquid. 
If the density of the electron liquid is not too small, it is enough to treat the electron-electron interaction in a relaxation time approximation.
In terms of the angular momentum components $h_l(y)$, the collision integral becomes,
\begin{align}
\mathcal{I}( h)&=-\frac{ v}{\ell_0}[
h^e_1(y)\cos\theta+ h^o_1(y)\sin\theta+ h^e_2(y)\cos2\theta
\notag\\&\quad
+ h^o_2(y)\sin2\theta
]-\frac{v }{\ell_{ee}}[
h^e_2(y)\cos2\theta+ h^o_2(y)\sin2\theta
].\label{collintegral}
\end{align}
This encodes that only impurity collisions (and not electron-electron interactions) relax momentum. 
Given the translation invariance along $x$, charge conservation $\partial_yJ_y=0$ enforces for the current in y-direction $h^o_1(y)=0$ for all $y$. 
Using length scales as coefficients, we can evaluate the kinetic equation for each moment separately,
\begin{align}
\tfrac{1}{2}\partial_y  h_2^o(y)-E_x
&=-\ell\inv_0 h^e_1(y)\label{eq:order1e}\\
\tfrac{1}{2}\partial_y  h_2^e(y)+\partial_yh_0
-R_c\inv  h_1^e+E_y&=0 \label{eq:order1o}\\
2R_c\inv  h_2^o&=-\ell\inv h^e_{2}(y)\\
\tfrac{1}{2}\partial_y h_1^e(y)-2R_c\inv  h_2^e&=-\ell\inv h^o_{2}(y).
\end{align}
Where the zero-field mean free path is $\ell=\ell_0\ell_{ee}/(\ell_0+\ell_{ee})$ and the cyclotron radius $R_c=v/\omega_c$. 
These four coupled equations need to be solved for $h^e_1$, $h^e_2$, $h^o_2$ and $E_y$. Note that they can be reformulated as the single Eq.~\ref{eq:hydro1} where the microscopic length scale is $\gamma=\sqrt{\ell\ell_0}$.

Under the simplifying assumption that the boundary scattering happens at a sharply defined hard wall, like a bounce, one can define a reflectivity $0<r_j<1$. The boundary conditions are then
\begin{align}
h_1^e(\pm w/2)&=r_j h_{1,bulk}^e,
\end{align}
which enforces the current density at the boundary to be at most as large as in the bulk without any boundaries. The bulk current density assumes the Drude value $h_{1,bulk}^e=\ell_0 E_x$. Neglecting the non-equilibrium density $h_0$, the solution for the longitudinal current profile is then
\begin{align}
     h^e_1(y)&=
    \ell_0 E_x \left(1-(1-r_j) 
    \frac{\cosh (2y/\ell_c)}
    {\cosh(w/\ell_c)}\right),
    \label{eq:h1esolution}
\shortintertext{with}
    \ell_c&=\sqrt{\frac{\ell \ell_0 R_c^2}{4 \ell^2+R_c^2}}.
\end{align}
The transverse electric field takes the form stated in Eq.~\eqref{eq:result2harm},
\begin{align}
    E_y&=R_c\inv  h_1^e+\tfrac{1}{2}\partial_y  h_2^e(y)\nonumber\\
    &=
    \frac{E_x}{R_c}
    \left(\ell_0-\left(2 \ell+\ell_0\right) (1-r_j) 
    \frac{\cosh (2y/\ell_c)}
    {\cosh(w/\ell_c)}
    \right),\label{eq:result2harmapp}
\end{align}
while the total longitudinal current and Hall resistance are given by
\begin{align}
    J_x&=
    \ell_0 E_x
    \left(w-(1-r_j)\ell_c 
    \tanh(w/\ell_c)\right)\nonumber\\
    \rho_{xy}&=
    \frac{1}{R_c} 
    \left(1-\frac{2 \ell (1-r_j) \ell_c 
    \tanh (w/\ell_c)}{\ell_0 \left(w-(1-r_j) \ell_c \tanh (w/\ell_c)
    \right)}\right).
    \label{eq:resultrhoxyapp}
\end{align}
As mentioned previously, in the Gurzhi-regime with $\ell_0\gg w$ and $\sqrt{\ell\ell_0}\gg w$, $E_y$ and $\rho_{xy}$ as given by Eqs.~(\ref{eq:result2harmapp},\ref{eq:resultrhoxyapp}) can change sign. To see this for the Hall resistivity, one can approximate at weak magnetic fields ($R_c\to\infty$) $\ell_c^2\approx\ell_0\ell>w^2$, yielding for an expansion in $\ell_c\to\infty$
\begin{align}
    \rho_{xy}&=\frac{1}{R_c} \left(1-\frac{6\ell^2}{w^2}\right).
\end{align}
A very similar result was obtained in~\cite{Scaffidi2017}.

\subsection{Near hydrodynamic corrections}
In the next step we elaborate on the statement that current density and stress at the boundary are generically determined by independent boundary conditions. Most notably, in ballistic transport and also the intermediate Gurzhi regime slip and stress should be considered independent.
To this end, we continue the mode expansion to third order, which only adds a small correction to the current profile and the transverse electric field in Eq.~\eqref{eq:result2harm}. Obviously, considering an additional angular moment allows to fix not only the boundary condition for the current but also for the stress. But as it turns out, the resulting boundary conditions can be formulated in terms of observable quantities. 

The coupled set of differential equations up to third order is given by
\begin{align}
\tfrac{1}{2}\partial_y  h_2^o
+E_x
&=-\ell_0\inv h^e_1
\label{eq:order1ethree}\\
\tfrac{1}{2}\partial_y  h_2^e-\omega_c  h_1^e+\partial_yh_0+E_y
&=0
\label{eq:order1othree}
\\
+\tfrac{1}{2}\partial_yh_3^o
+2\omega_c  h_2^o
&=-\ell\inv h^e_{2}\\
\tfrac{1}{2}\partial_y h_1^e
-\tfrac{1}{2}\partial_yh_3^e
-2\omega_c  h_2^e
&=-\ell\inv h^o_{2}\\
-\tfrac{1}{2}\partial_y h_2^o
+3\omega_c  h_3^o
&=-\ell\inv h^e_{3}\\
\tfrac{1}{2}\partial_y h_2^e
-3\omega_c  h_3^e
&=-\ell\inv h^o_{3}.
\end{align}
This can be solved explicitly, but results in fairly long expressions.
Given symmetric boundaries, it follows that even harmonics are odd functions of $y$ and odd harmonics are even in $y$.
Evaluating Eqs.~\eqref{eq:order1ethree} and \eqref{eq:order1othree}, it is
\begin{align}
\tfrac{1}{2}\partial_y  h_2^o(\pm w/2)
&=-\ell_0\inv h_1^e(\pm w/2)+E_x
\label{eq:thorder1}\\
\tfrac{1}{2}\partial_y  h_2^e(\pm w/2)&=R_c\inv h_1^e(\pm w/2)-E_y(\pm w/2).
\label{eq:thorder2}
\end{align}
It is more convenient integrate the $y$ dependence and express the stress at the boundary in terms of the total current $J_x$ and the Hall voltage $U_{H}$,
\begin{align}
    \frac{h_2^o(+w/2)- h_2^o(-w/2)}{2}-wE_x=-\ell_0\inv J_{x}\nonumber\\
    \frac{ h_2^e(+w/2)- h_2^e(-w/2)}{2}+U_{H}=R_c\inv J_{x}.
\end{align}
This clarifies the dependence of the stresses at the boundary on the transverse electric field. After dividing by $J_{x}$, one can write them in terms of longitudinal $\rho_{xx}$ and Hall $\rho_{xy}$ resistance, ($L$ is the length of the channel) 
\begin{align}
    \frac{h_2^o(w/2)}{J_x}-\rho_{xx} w/L=-\ell_0\inv
    \label{eq>stress1app}\\
    \frac{h_2^e(w/2)}{J_x}+\rho_{xy}=R_c\inv.
    \label{eq>stress2app}
\end{align}
Here, $L$ is the length of the channel, and we used $h_2^o(+w/2)=-h_2^o(-w/2)$ and the same for $h_2^e$. To determine the stress parameters at the boundary, it is therefore enough to measure the cyclotron radius $R_c$, the disorder mean free path $\ell_0$ and the resistivity matrix. 
As mentioned, the expansion to third order is analogous to an extended hydrodynamic approach where the stress at the boundary is fixed independently from current density. In the usual approach, corresponding to a second order expansion, both quantities are interdependent by means of the phenomenological slip length $\ell_s$~(cf.~\cite{Torre2015}),
\begin{align}
    \ell_s&=-\frac{h_1^e(w/2)}{\partial_y h_1^e(y)|_{w/2}}=\frac{\ell_c r_j}{2(1-r_j)\tanh(w/\ell_c)}.
\end{align}
This represents a single boundary condition which encapsulates non-local correlations with the common length scale $\ell_s$. In contrast, in Eqs.(\ref{eq>stress1app},\ref{eq>stress2app}) both Fermi surface deformations $h_2^e$ and $h_2^o$ are fixed independently.

\section{Kinetic equation: General formalism}
\label{sec:four}
The solution of the kinetic equation forms the basis for our main results. We obtain analytical solutions for the limiting cases with $\ell_{ee}\to\infty$ and either $R_c\to\infty$ or $\ell_0\to \infty$. The general case is solved numerically.
Starting from the kinetic equation, Eq.~\eqref{eq:kinetic} it is first necessary to approximate the collision integral. The moment expansion (Eq.~\eqref{collintegral}) provides an intuitive picture how to do this in the relaxation time approximation. Using the angular momentum projections, the kinetic equation can be re-expressed identically as [cf. Eqs.~(\ref{eq:main1},\ref{eq:main2})]
\begin{align}
&R_c\inv\partial_\theta h(y,\theta)
+\sin\theta\partial_y h(y,\theta)
=-\frac{ h(y,\theta)}{\ell}+
F(y,\theta)
\label{eq:general1}
\end{align}
with the inhomogeneous term
\begin{align}
F(y,\theta)
&=E_x\cos\theta-
E_y\sin\theta
\notag\\
&+\frac{h_0(y)}{\ell}
+\frac{ h_1^e(y)\cos\theta+h_1^o(y)\sin\theta}{\ell_{ee}}
\label{eq:general2}
\end{align}
Here, the collision integral $h_0(y)/\ell$ contains all angular modes in the relaxation time approximation. To ensure charge conservation and conversation of momentum for e-e collisions, $h_0$ and $h_1$ are then subtracted again, creating the additional inhomogeneities in Eq.~\eqref{eq:general2}.

Following ref.~\cite{deJong1995}, we first consider a classical bounce trajectory. Then the following relations hold at the boundary,
\begin{align}
h(-w/2,-|\theta|)&=r_h(\theta)h(-w/2,|\theta|)
\notag\\&
+\int_{0}^{\pi}\frac{\dd \theta'}{\pi}  (1-r_h(\theta'))h(-w/2,\theta')
\notag\\
h(w/2,|\theta|)&=r_h(\theta)h(w/2,-|\theta|)
\notag\\&
+\int_{-\pi}^{0}\frac{\dd \theta'}{\pi}(1-r_h(\theta'))
 h(w/2,\theta')
\label{eq:general3}
\end{align}
Compared to the mode expansion, this boundary condition does not fix current or stress, but stipulates that a bounce will either be specular with probability $r_h$ or fully diffusive with probability $1-r_h$, where the specularity parameter $r_h$ and the no-slip parameter $r_j$ from the moment expansion are not simply related.

Importantly, the identification of positive and negative angles in these boundary conditions means that a solution connects periodically only after two times the channel width. Equations~(\ref{eq:general1},\ref{eq:general2},\ref{eq:general3}) determine the  flow of a classical electron liquid with short range interactions. Additionally, the charge distribution of both sample and environment should be calculated using Gauss's law. We separate the latter step by demanding self-consistently that 
\begin{align}
E_y=\tfrac{1}{R_c}h_1^e(y)+\tfrac{1}{2}\partial_y h_2^e-\partial_yh_0,\label{eq:selfconsistentEyapp}
\end{align} 
which leads to a solution of the Boltzmann equation where sources proportional to $\sin\theta$ are automatically included, irrespective of the actual solution obtained for $E_y$. This allows to determine the longitudinal current without needing to specify the dielectric environment and $E_y$, which we postpone to a later stage. We note that such an approach works as long as the current density is not dramatically altered by charging effects. We treat effects from unscreened charges in App.~\ref{sec:A}.

\subsection{Non-interacting flow}
In the limit $\omega_c=0$ and $\ell_{ee}=\infty$ the solution for $h(y,\theta)$ satisfying Eq.~\eqref{eq:general1} can be given in closed form~\cite{Alekseev2018}. The solution can also be reconstructed from the ansatz by~\citeauthor{deJong1995}~\cite{deJong1995} for an interacting system, but this was not done at the time. Also, their ansatz does not generalize to include magnetic fields. 
To find the solution, we recur to the method of characteristics. This method allows us to calculate the distribution function by tracing a \emph{flow characteristic} curve that is defined at the boundary. Introducing $\xi$ to parametrize the boundary and $t$ to trace the characteristics, the boundary is defined as $y(0,\xi)=-w/2$ and $\theta(0,\xi)=\xi$. Then the characteristics are determined by the relations
\begin{align}
\frac{\partial y}{\partial t}&=
\sin\theta &y(t,\xi)&=\sin\xi t-w/2\nonumber\\
\frac{\partial \theta}{\partial t}&=
0&\theta(t,\xi)&=\xi.
\end{align}
These characteristics correspond to the classical trajectories of the electrons in the absence of scattering, either due to interactions or impurities. 

Without a magnetic field, $h_0(y)=0$, and the boundary conditions for the distribution function $\bar h(t,\xi)$ written in terms of $t$ and $\xi$ become
\begin{align}
\bar h(0,|\xi|)&=r_h\bar h(0,-|\xi|) 
\notag\\
\bar h(-t_1,-|\xi|)&=r_h\bar h(t_1,|\xi|) 
\notag\\
t_1 &=w/\sin|\xi|
\end{align}
It is now possible to find the form of the distribution function,
\begin{align}
    \bar h(t,\xi)
    &=e^{-\frac{t}{\ell_0}}c(\xi)+\bar h_{in}(t,\xi)
\shortintertext{with}
    \bar h_{in}(t,\xi)
    &=\int_0^t e^{\frac{t'-t}{\ell_0}}\bar F(t',\xi)\mathrm{d}t'
\end{align}
yielding for the boundary conditions
\begin{align}
    c(|\xi|)&=r_h c(-|\xi|)\\
    e^{\frac{t_1}{\ell_0}}c(-|\xi|)+
    \bar h_{in}(-t_1,-|\xi|)
    &=\notag\\
    r_h(
    e^{-\frac{t_1}{\ell_0}}c(|\xi|)+&
    \bar h_{in}(t_1,|\xi|)).
\end{align}
The solution is
\begin{align}
    c(|\xi|)&=
    \frac{r_he^{\frac{t_1}{\ell_0}}}
    {e^{\frac{2t_1}{\ell_0}}-r_h^2}
    (
    r_h\bar h_{in}(t_1,|\xi|)-
    \bar h_{in}(-t_1,-|\xi|)
    ),
\end{align}
which vanishes in the case of fully diffusive boundaries, $r_h=0$. With $F(t,\xi)=eE_x\cos\xi$ one obtains
\begin{align}
    \bar h_{in}(t,|\xi|)&=
    eE_x\cos(\xi)
    \left(
    e^\frac{t}{\ell_0}-1
    \right).
\end{align}
Putting everything together, the result reads for $\theta>0$
\begin{align}
\bar h(t,|\xi|)&=
e E_x \ell_0\cos (\xi ) \left(1-\frac{(1-r_h)e^{-\frac{t}{\ell_0}} }{1-r_he^{-\frac{t_1}{\ell_0}}}\right)
\end{align}
and analogously for $\theta<0$:
\begin{align}
\bar h(-|t|,-|\xi|)&=
    e E_x \ell_0 \cos (|\xi|) \left(1-\frac{(1-r_h) e^{\frac{|t|}{\ell_0}}}{e^{\frac{t_1}{\ell_0}}-r_h}\right)
\end{align}
After transforming back, both solutions fall together to
\begin{align}
     h(y,\theta)&=
    e E_x \ell_0 \cos (\theta ) 
    \left(1-(1-r_h)\frac{e^{\frac{(w-2 y \sgn{\theta}) }{2 \sin|\theta| \ell_0}}}{e^{\frac{w}{\sin|\theta|\ell_0}}-r_h}\right).
    \label{eq:nonintsolution}
\end{align}
In this expression, $r_h$ can be taken as dependent on the angle of incidence. 

\begin{figure}
    \centering
    \includegraphics[width=.9\columnwidth]{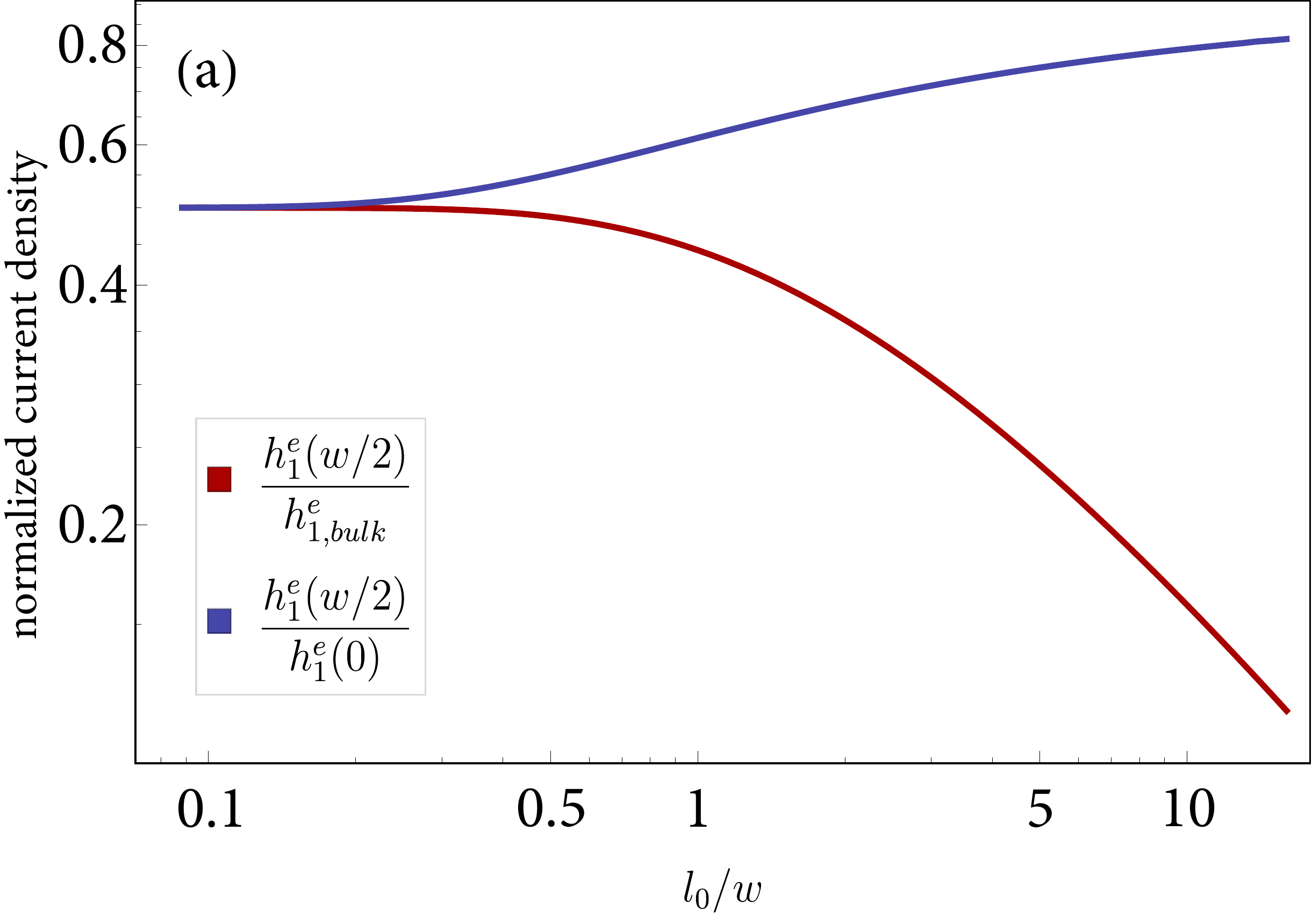}

    \includegraphics[width=.9\columnwidth]{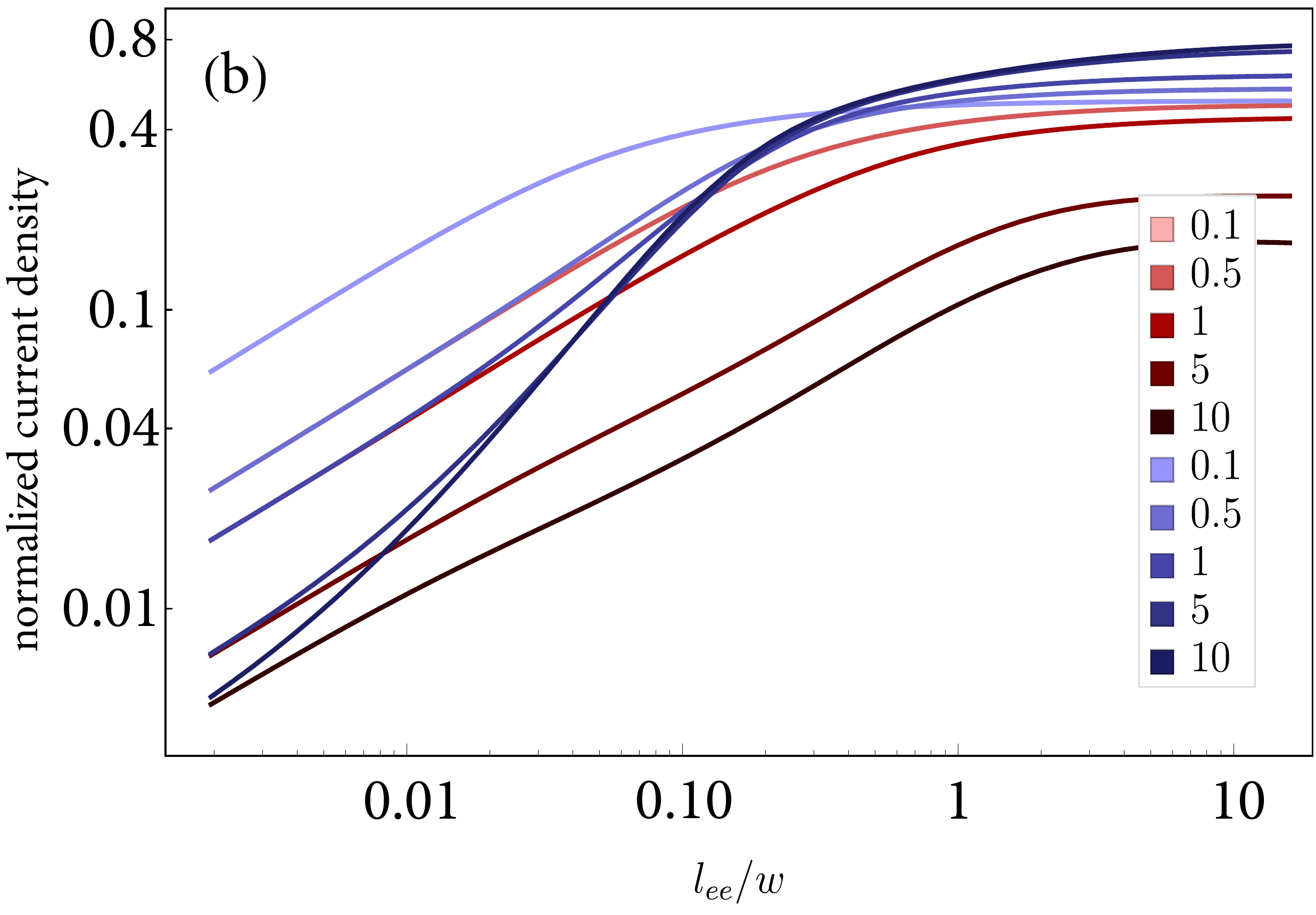}
    \caption{(a) Current density at the boundary for reflectivity $r_h=0$ and without interactions. The current is normalized either with respect to the bulk current density (red) or the current density in the center of the channel (blue). (b) Current density at the boundary as a function of $\ell_{ee}$ for different disorder mean free paths $\ell_0\in \{0.1,0.5,1,5,10\}$. The ratio boundary/bulk current density is roughly $\sqrt{\ell_{ee}/\ell_0}$. Colors like beforehand and $r_h=0$.}
    \label{fig:mincurrent}
\end{figure}
Using this solution, it is easy to check how small the current at the boundaries can be. Fig.~\ref{fig:mincurrent} shows the current density at the boundary as a function of $l_0$ for $r_h=0$. In the ohmic limit ($l_0\to 0$), the current in the center of the channel is equal to its bulk value. At the boundary, the value is half of that. This is due to the nature of the boundary, which for $r_h=0$ fully dissipates the momentum of impacting electrons, while it is inconsequential for the momentum of the incoming particle before it hits the wall. The current density is therefore half of its bulk value, with the entire current being transported by incoming particles before their bounce.
In the opposite limit $l_0\to\infty$, the current at the boundary increases only logarithmically slowly, which matches the behavior of the current in the middle, resulting in an asymptotically flat profile.
Obviously, hard wall bounces cannot reduce the current at the boundary to zero, which would correspond to no-slip boundary conditions. Instead we observe that $h_1^e(w/2)/h_1^e(0)\geq 1/2$. 

While the current density at the boundary will indeed asymptotically approach zero for $l_0\to 0$ once electron-electron interactions are included (Fig.~\ref{fig:mincurrent}, cf.~\cite{deJong1995}), we caution that no-slip or nearly no-slip boundary conditions still represent an extreme case which can result in issues like the one we discussed in connection with the Hall resistivity. Given the slow decrease of the residual current at the boundary, we conclude that no-slip boundary conditions only apply deep in the hydrodynamic regime. To be precise, inserting the hydrodynamic result for the current [Eq.~\eqref{eq:h1esolution}] into the kinetic equation for $r_h=0$ [Eq.~\eqref{eq:general1}] and integrating for the longitudinal current component yields for the coefficient $r_j$ the self-consistency condition
\begin{align}
    r_j&\approx\frac{r_j}{2}+\frac{2}{3\pi}\sqrt{\frac{\ell_{ee}}{\ell_0}}(1-r_j)
    \approx \frac{4}{3\pi}\sqrt{\frac{\ell_{ee}}{\ell_0}}+\mathcal{O}\left(\frac{\ell_{ee}}{\ell_0}\right),
\end{align}
which is valid for $\ell_{ee} \ll \ell_0 \ll w$.

\subsection{Weak magnetic fields}
\label{sec:characteristics}
The solution including a magnetic field proceeds in a similar fashion as detailed for the non-interacting case. 
The characteristics are now parametrized as
\begin{align}
\frac{\partial y}{\partial t}&=
\sin\theta &y(t,\xi)&=\left(\cos\xi-\cos(R_c\inv t+\xi)\right)R_c-w/2\\
\frac{\partial \theta}{\partial t}&=
R_c\inv&\theta(t,\xi)&=R_c\inv t+\xi
\end{align}
This leads to several regions where the solutions are independently defined. As long as $|(y\pm w/2)R_c\inv+\cos\theta|<1$, very similar boundary conditions hold as detailed in non-interacting case, with the generalized definitions
\begin{align}
    t&=(\theta-\sgn{\theta}\arccos((y+w/2)R_c\inv+\cos\theta))R_c
    \label{eq:deft}\\
    \xi&=\sgn{\theta}\arccos((y+w/2)R_c\inv+\cos\theta)
    \label{eq:defxi}\\
    t_1&=(\arccos(\cos\xi-wR_c\inv)-|\xi|)R_c
\end{align}
We also allow for nonzero density on the boundary, $h_+\equiv h_0(w/2)$ and $h_-\equiv h_0(-w/2)$. The solution is then
\begin{widetext}
\begin{align}
	\bar h(t,|\xi|)&=\frac{e^{-\frac{t}{\ell}}}{e^{\frac{t_1}{\ell}}-R^2e^{-\frac{t_1}{\ell}}}
	\left((1-r_h)h_-e^{\frac{t_1}{\ell}}+(1-r_h)r_hh_++R^2\bar h_{in}(t_1,|\xi|)-r_h\bar h_{in}(-t_1,-|\xi|)\right)+\bar h_{in}(t,|\xi|)\\
	\bar h(t,-|\xi|)&=\frac{e^{-\frac{t}{\ell}}}{e^{\frac{t_1}{\ell}}-r_h^2e^{-\frac{t_1}{\ell}}}
	\left((1-r_h)r_hh_-e^{-\frac{t_1}{\ell}}+(1-r_h)h_++r_h\bar h_{in}(t_1,|\xi|)-\bar h_{in}(-t_1,-|\xi|)\right)+\bar h_{in}(t,-|\xi|)
\end{align}
\end{widetext}
In case that $|(y+w/2)R_c\inv+\cos\theta|<1$ and $|(y-w/2)R_c\inv+\cos\theta|>1$, the characteristics do not reach the upper boundary, yielding with the same boundary conditions instead ($t_2=(2\pi-2|\xi|)R_c$)
\begin{align}
\bar h(0,|\xi|)&=r_h h(t_2,-|\xi|)\\
\bar h(t,|\xi|)&=e^{-\frac{t}{\ell}}r_h\frac{\bar h_{in}(t_2,|\xi|)+h_-(1-r_h)}{1-r_he^{-\frac{t_2}{\ell}}}+\bar h_{in}(t,|\xi|)\\
\bar h(t,-|\xi|)&=e^{-\frac{t_2-t}{\ell}}r_h\frac{\bar h_{in}(t_2,|\xi|)+h_-(1-r_h)}{1-r_he^{-\frac{t_2}{\ell}}}
\notag\\&\quad
+\bar h_{in}(t_2-t,|\xi|)
\end{align}
If $|(y+w/2)R_c\inv+\cos\theta|>1$ and $|(y-w/2)R_c\inv+\cos\theta|<1$, characteristics originating from the upper boundary will not reach the lower boundary, leading to the replacement $w\rightarrow-w$ in Eqs.~(\ref{eq:deft},\ref{eq:defxi}) and $t_3=-2|\xi| R_c$. Given these changes, we then obtain
\begin{align}
r_h\bar h(0,|\xi|)&= h(t_3,-|\xi|)\\
\bar h(t,|\xi|)&=e^{-\frac{t}{\ell}}\frac{\bar h_{in}(t_3,|\xi|)-h_+(1-r_h)}{r_h-e^{-\frac{t_3}{\ell}}}
\notag\\&\quad
+\bar h_{in}(t,|\xi|)\\
\bar h(t,-|\xi|)&=e^{-\frac{t_3-t}{\ell}}\frac{\bar h_{in}(t_3,|\xi|)-h_+(1-r_h)}{r_h-e^{-\frac{t_3}{\ell}}}
\notag\\&\quad
+\bar h_{in}(t_3-t,|\xi|)
\end{align}
Finally, if there exists a region with  $|(y\pm w/2)R_c\inv+\cos\theta|>1$, the boundary does not uniquely define the flow in this region. This happens if $w /R_c>2$. Away from the boundaries, $ h(y,\theta)$ is assumed to be continuous in $\theta$. Note that directly at the boundary some discontinuities are expected due to scattering losses ($1>r_h=const.$).
The parametrization is in this region is given by
\begin{align}
    a&=y+\cos\theta R_c\\
    b&=\theta R_c
\end{align}
The solution reads
\begin{align}
    \tilde h(a,b)&=
    e^{-\frac{b}{\ell}}\frac{\tilde h_{in}(a,\pi R_c)-\tilde h_{in}(a,-\pi R_c)}{2\sinh(\pi R_c/\ell)}
    +\tilde h_{in}(a,b).
\end{align}
We use these parameterizations to solve the kinetic equation self-consistently. In the following we discuss the distribution function itself. This makes it possible to carefully assess the effects of various boundary conditions. The analytical solution in the ultraballistic limit but at finite magnetic field can be found in App.~\ref{sec:B}.

\subsection{Distribution function}
\begin{figure}
\includegraphics[width=.66\columnwidth]{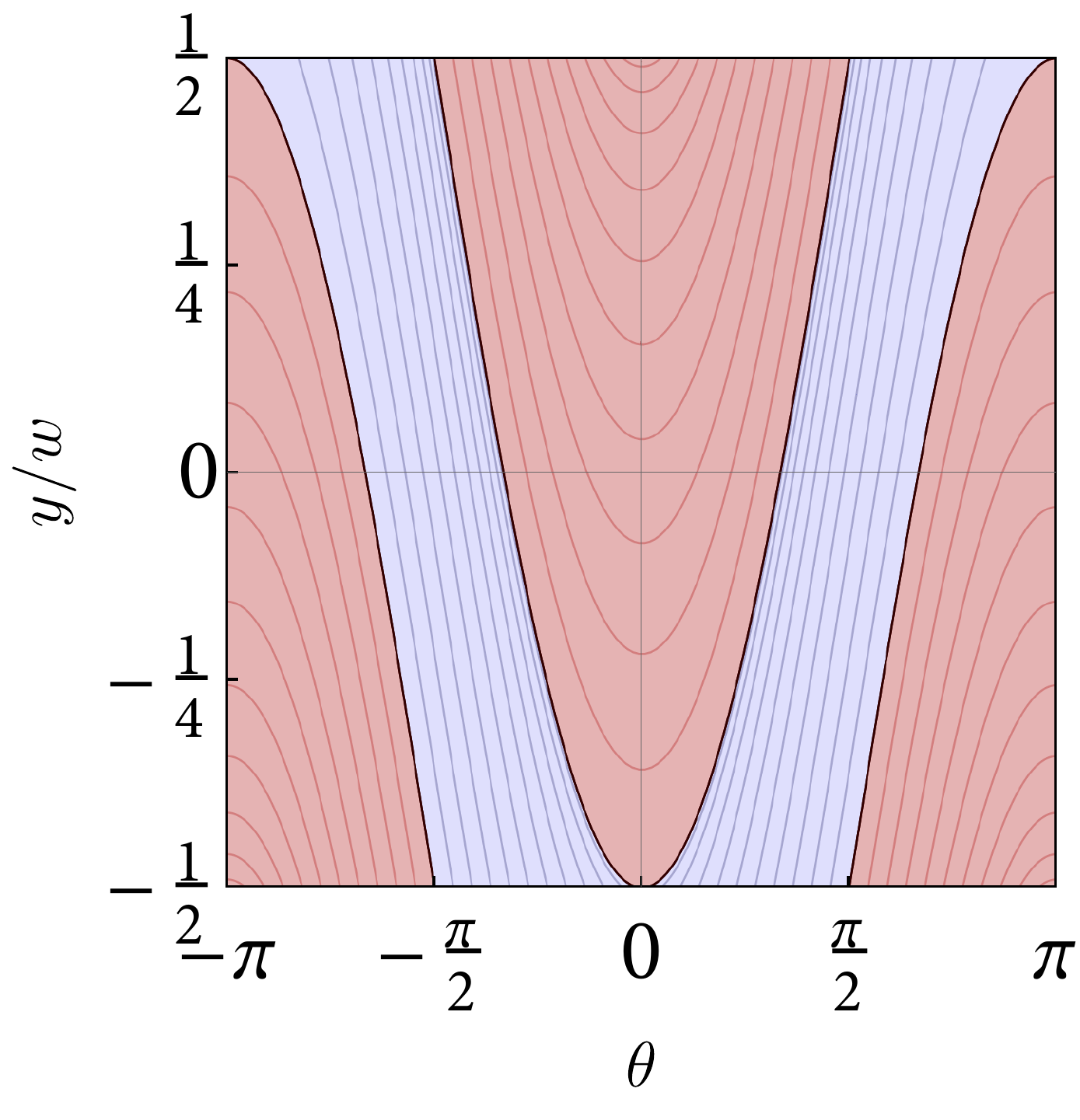}

\includegraphics[width=.66\columnwidth]{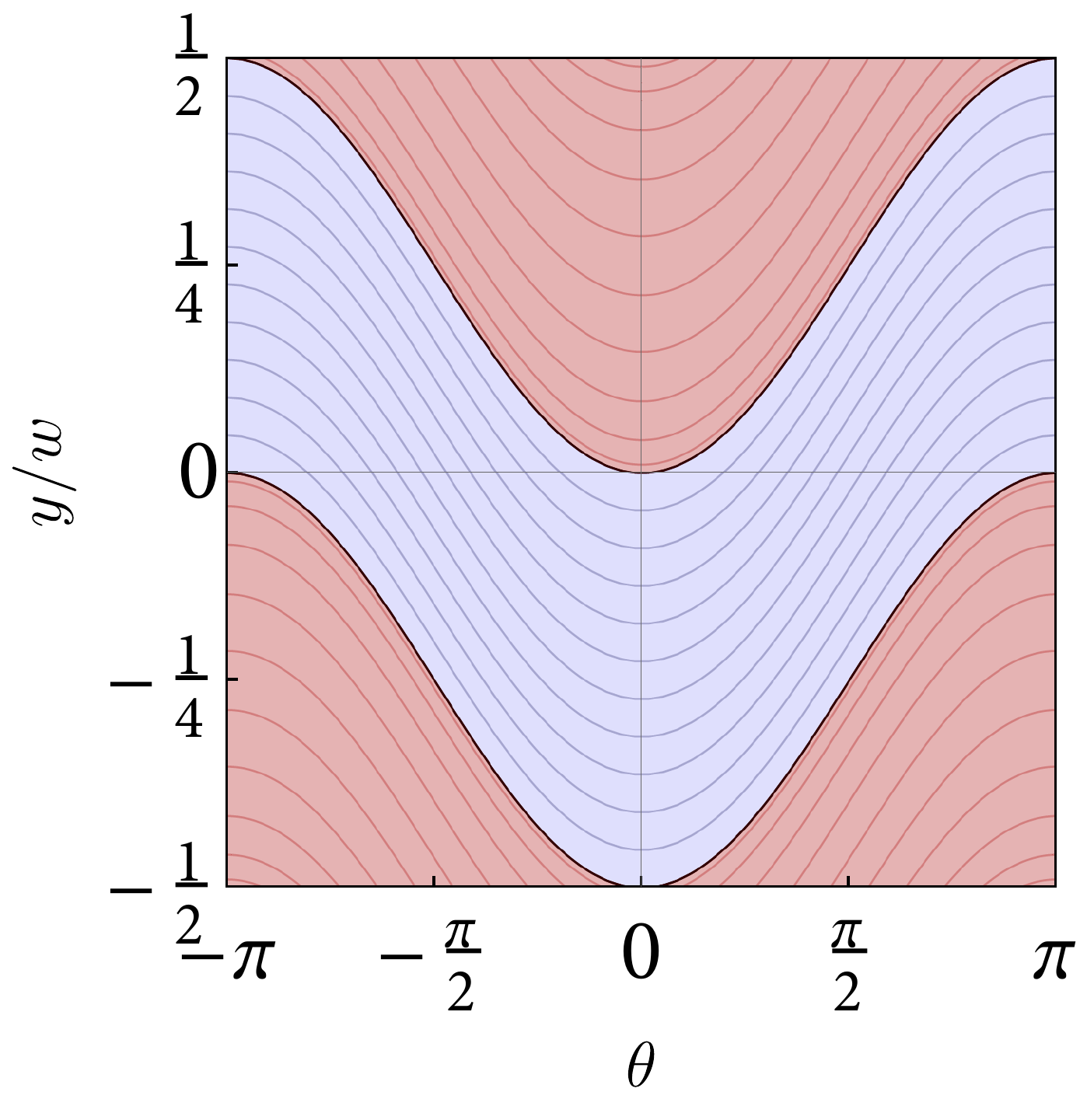}
\caption{Form of the characteristics of Eq.~\eqref{eq:general1} in the $y-\theta$ plane for $w/R_c=1$ (a) and $w/R_c=4$ (b). Shaded red are regions were the boundary conditions connect angles $\theta$ and $-\theta$ on the same side of the channel. Shaded in blue are regions which connect points periodically either along $y$ according to $(w/2,\pi-\theta)\to (-w/2,\theta)\to (-w/2,-\theta)\to (w/2,-\pi+\theta)$ in (a) or along $\theta$ according to $(y,-\pi)\to (y,\pi)$ in (b). }
\label{fig:characters}
\end{figure}

Regarding the distribution function in the $y$-$\theta$ plane, the primary effect of a magnetic field is to bend the flow characteristics of the kinetic equation (Fig.~\ref{fig:characters}). 
\begin{figure}
\includegraphics[width=.9\columnwidth]{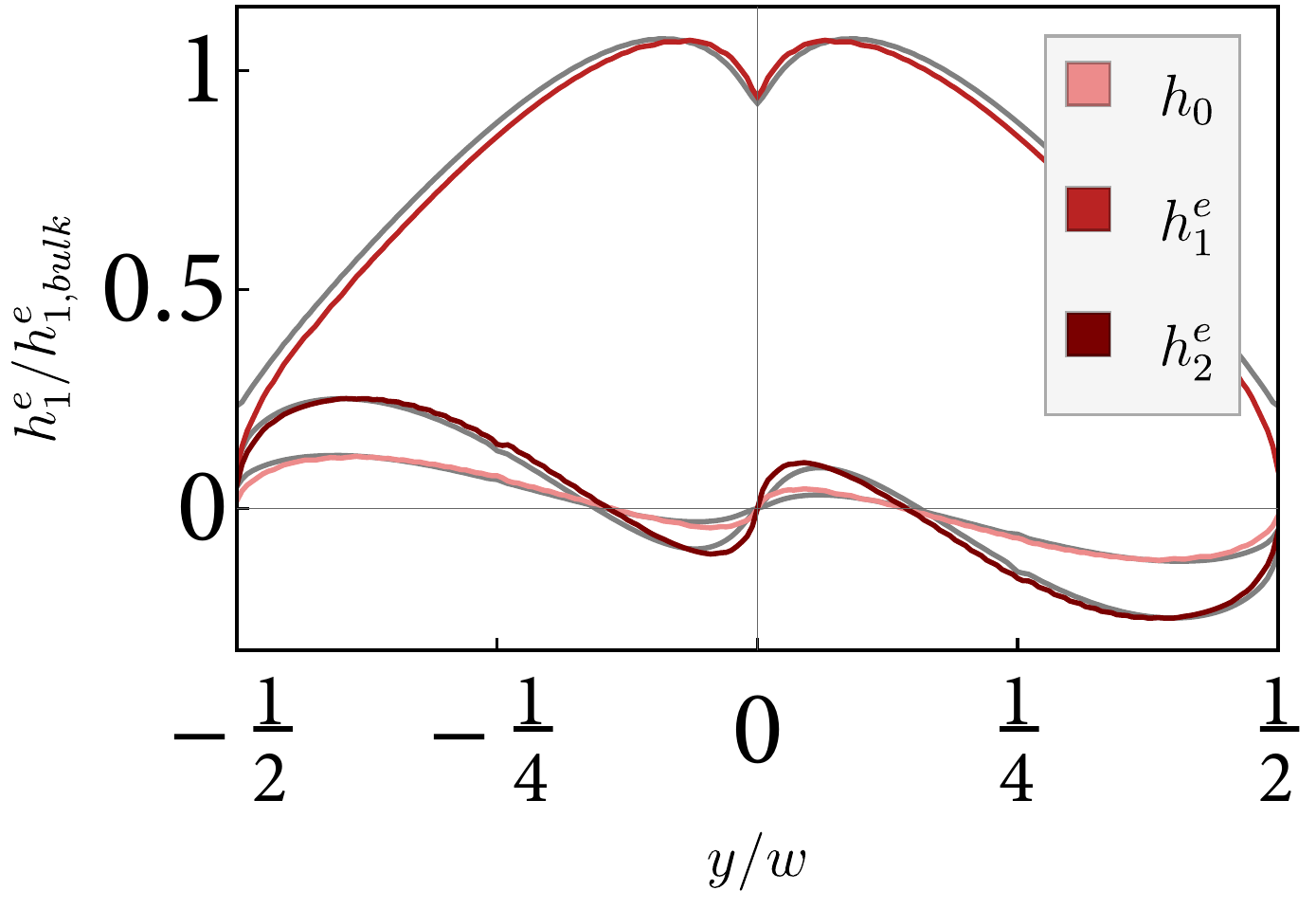}

\includegraphics[width=.9\columnwidth]{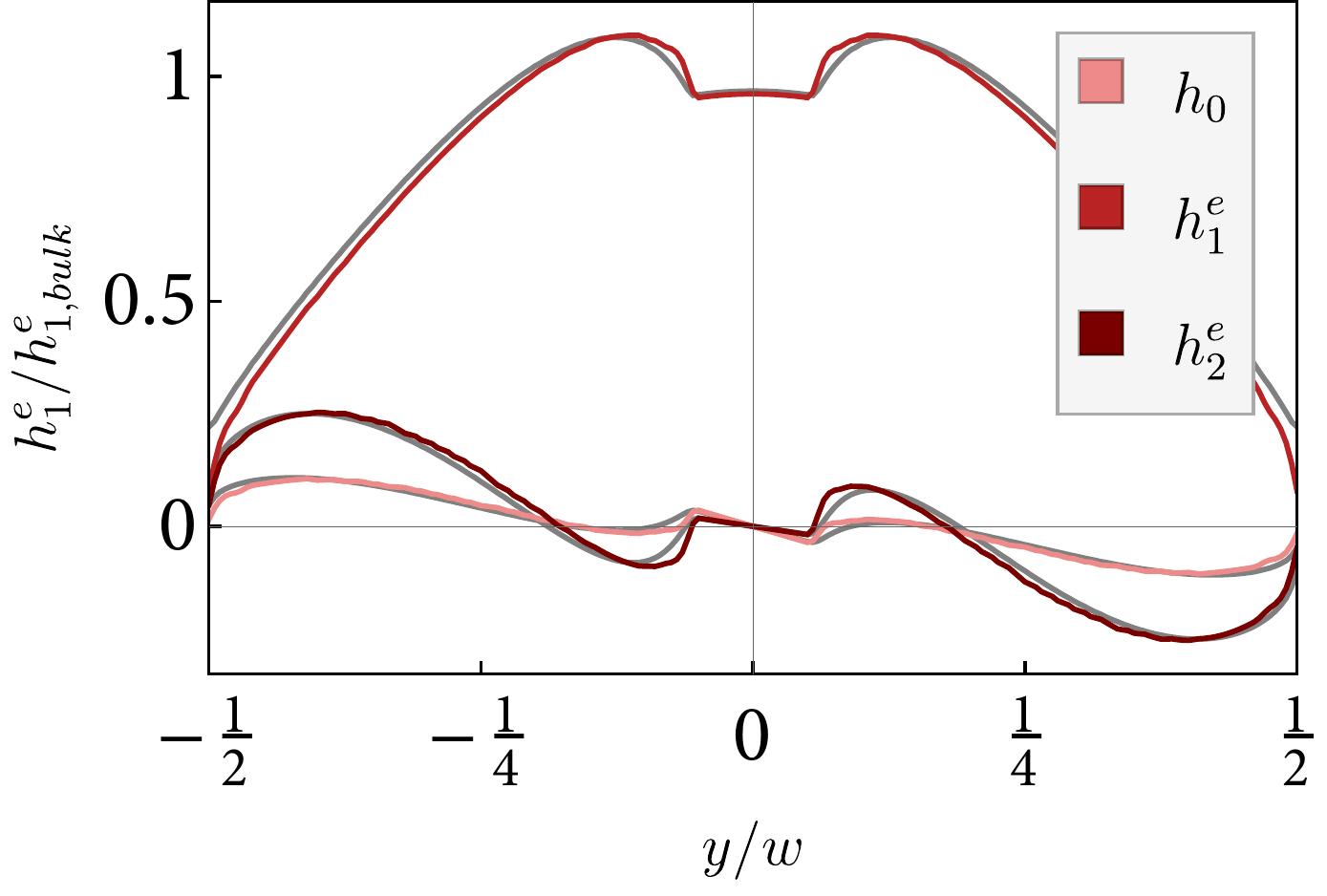}
\caption{Signatures of hard wall boundary conditions. In a magnetic field fulfilling $w/R_c\geq 4$ the current profile develops kinks, with similar features appearing in the non-equilibrium density and the stress. Both completely diffusive boundaries (red) and a model with specular anisotropy $\alpha=5$ (gray) lead to very similar results. The values are $\ell_0=3w$, $\ell_{ee}=7w$, $r_h=0$ and $w/R_c=4.0$ (a) and $w/R_c=4.5$ (b). Note that the solution for the charge density does not contain the correction due to electrostatic effects, which is minute in most cases.}
\label{fig:soffer}
\end{figure}
In the presence of a magnetic field some characteristic lines originating from the boundaries at $y=\pm w/2$ do no longer connect with the opposite boundary but instead with the same side of the channel at the sign-reversed angle. If $w/R_c>2$ no characteristic will hit the other boundary, resulting in true bulk behavior in the middle of the sample.
This sounds very much like a description of cyclotron orbits of quasiparticles in the liquid. However, the characteristics of the kinetic equation and quasiparticle trajectories are physically distinct, the former applies for the distribution function also in the presence of interactions and disorder scattering, cyclotron orbits on the other hand can only be completed for a very long mean free path and do not correspond to the trajectory which quasiparticles follow in an interacting material. 

Since the magnetic field breaks time reversal symmetry it is no longer true that the distribution function at positive $\theta$ is simply related to its value at $-\theta$. 
In the same vain, the steady state may now include a non-equilibrium density $h_0$. 
Fortunately, not all these effects are equally important.
The finite current along the channel is a direct result of the applied electric field $E_x$ and is normally the largest contribution in
Eq.~(\ref{eq:general1}) to $E_y$, Eq. \eqref{eq:selfconsistentEyapp}. The $h_0$ and $h_2^e$ terms correspond to secondary effects originating from the unequal current distribution in $h_1^e$, and can often be neglected without affecting the resulting distribution function qualitatively. They do however have a quantitative influence on observables like the magnetoresistance. 

These expectations are confirmed by a numerical solution of Eqs.~(\ref{eq:general1},\ref{eq:general2},\ref{eq:general3}), shown exemplary in Fig.~\ref{fig:soffer} for ballistic transport and two choices of boundary specularity,
\begin{align}
r_h^{(1)}(\theta)&=const.\\
r_h^{(2)}(\theta)&=\exp(-\alpha \sin^2\theta).
\end{align}
The latter type of boundary condition was suggested in a microscopic calculation~\cite{Soffer1967} and corresponds to grazing trajectories being more specular than large angle bounces. We observe that a number of distinct cusps develop for $w/R_c>2$. 

\begin{figure}
\includegraphics[width=1.\columnwidth]{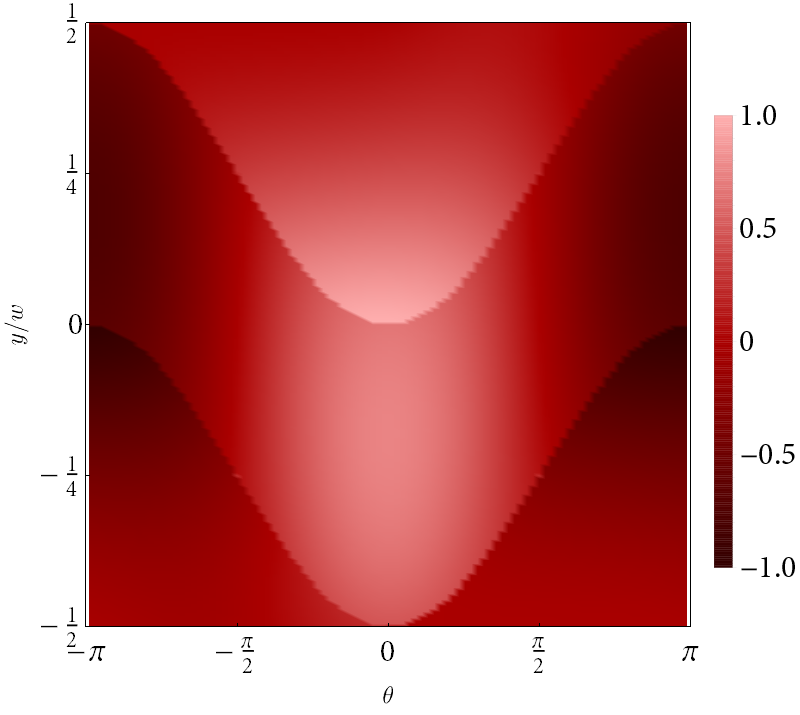}
\caption{
Example of the non-equilibrium distribution function $h(y,\theta)$ calculated from Eq.~\eqref{eq:general1}. The parameters are $\ell_0=3w$, $\ell_{ee}=7w$, $r_h=0.5$ and $w/R_c=4$. The regions where the boundary conditions have a large impact can easily be distinguished from the bulk part, which is only slightly altered due to the nonuniform current profile.}
\label{fig:hardbound}
\end{figure}

These features can be best explained by temporarily misidentifying the characteristics with particle trajectories, which in reality is only justified for $\ell_0\ll w$.
A typical distribution function is shown in Fig.~\ref{fig:hardbound}. First of all, in the ballistic regime the distribution function has pronounced discontinuities between bouncing trajectories and bulk trajectories. This is due to the long mean free path in the system, which essentially allows the bulk to decouple from the bouncing trajectories. The result of these discontinuities is a characteristic dip in the current profile where the boundary trajectories terminate. 

Importantly, if the boundary condition connects smoothly (example $r_h^{(2)}$), this sharp feature in the current profile is smoothened only slightly (Fig.~\ref{fig:soffer}). Indeed, for $w/R_c\geq 4$ and as long as it is true that $r_h(\theta)<1$, the central dip at $y=0$ is very robust.
When going from ballistic to hydrodynamic transport, the discontinuities in the distribution function will decay quickly when going from the boundary into the bulk due to the shorter effective mean free path, but they will not vanish completely. 

Secondly, with or without magnetic field any condition on the specularity of the wall can only ever determine how much current is lost at the boundary. It will not enforce the current to be zero, which would correspond to a no-slip condition for the current density. In particular, it is impossible to demand both that the distribution function is almost symmetric under reflection, $h(y,+\theta)=h(y,-\theta)$ and that the wall dissipates momentum. These two properties come together when transport is mostly hydrodynamic such that the mean free path due to electron-electron interactions is shorter than the width of the boundary layer. In this case, the electrons are slowed down due to interactions with electrons rebounding from the wall before actually hitting the wall.

Less important in the present context, we note that the distribution function $h(y,\theta)$ exhibits the suppression of backscattering at the respective boundaries, meaning that forward traveling ($\theta=0$) or backwards traveling ($\theta=\pi$) densities each vanish on one of the two boundaries. This is the precursor of chiral edge transport.

To summarize, for ballistic transport and in a magnetic field the distribution function will develop discontinuities which produce kinks in charge, current and stress profiles. As pointed out previously, this also implies spikes in the transverse electric field which cannot be removed just by slightly altering the diffusive boundary condition. 

 \begin{figure}
     \centering
     \includegraphics[width=1.\columnwidth]{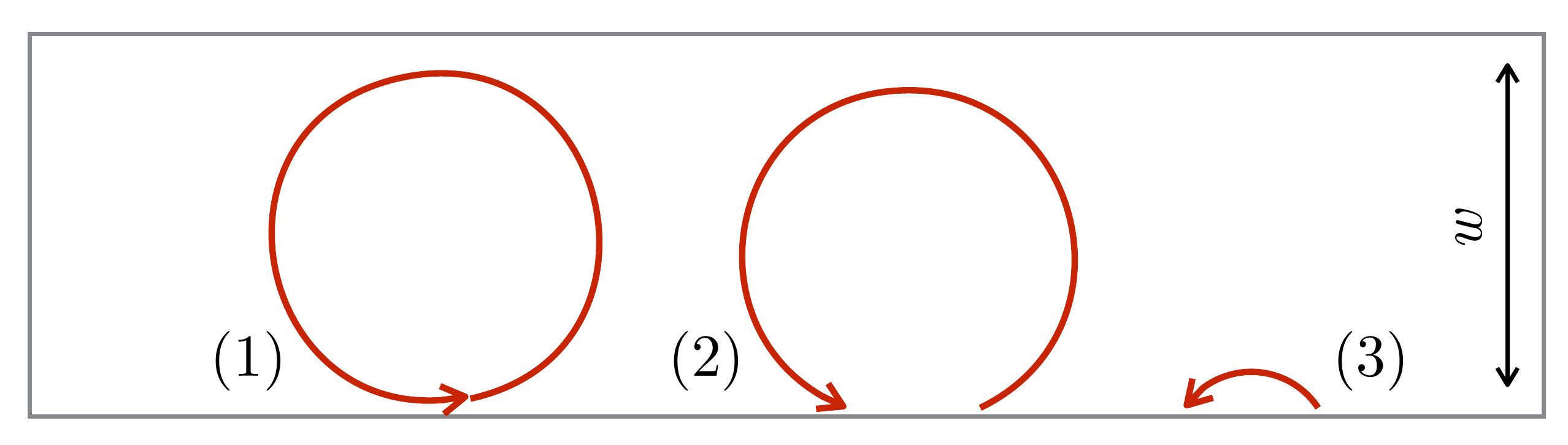}
     \caption{Ballistic trajectories which are sensitive to the chosen boundary properties for the channel.}
     \label{fig:schemeapp}
 \end{figure}

\subsection{Alternative boundary condition}
Guided by a semiclassical argument, we suggest an alternative approach to boundary scattering and the expected profile of $E_y$ and $h_1$. 
To this end, compare three particle trajectories close to the boundary, see Fig.~\ref{fig:schemeapp}. For hard wall bounces, trajectories $2$ and $3$ will be slowed down a lot while $1$ can complete many orbits unhindered. If on the other hand $1$ and $3$ are not completely non-interacting the momentum loss at the wall suffered by $3$ will also transpire a little to $1$. This effect is accounted for in the solution of the kinetic equation. However, a smooth boundary will couple trajectories $1$ and $2$ even without any electron-electron interaction, because it becomes possible for $2$ to not collide with the wall and vice versa, i.~e. the distinction between both is blurred. We note that this effect is not related to the boundary layer, which is also present and already contained in the kinetic equation via the stress tensor. Instead, a finite size boundary can be seen as a finite smoothness of the confining potential of the electron liquid. Without a magnetic field this detail is not important because almost all proximate trajectories will also collide with the wall in close proximity to each other. The only exception to this rule would be quasiparticles traveling near to and in parallel to the wall, but they have negligible weight in a long channel.

\begin{figure}
\includegraphics[width=.9\columnwidth]{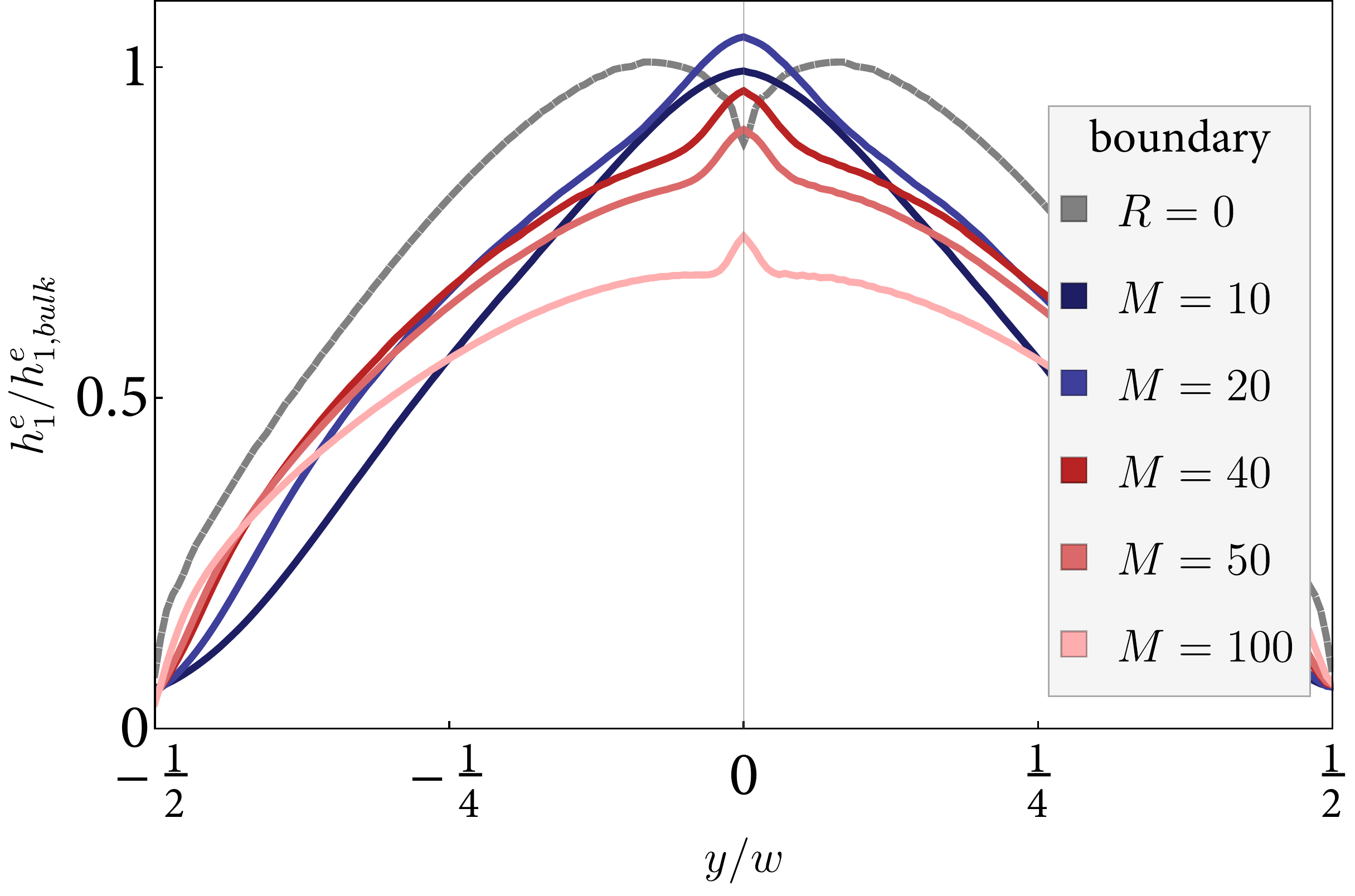}
\caption{A comparison of different current profiles of billiard-ball type hard wall bundaries (grey) with smooth confining potentials of varying degrees of smoothness $M\inv$ (colored). In the presence of a magnetic field qualitative differences emerge. Values $\ell_0=3w$, $\ell_{ee}=7w$, $R=0$ and $w/R_c=4$.  }
\label{fig:pathology}
\end{figure}

We thus test a model for the boundary which has a smooth and narrow absorptive layer where locally the effective mean free path is greatly reduced, $\ell_{\text{eff}}\inv(y)=\ell\inv+b(y)$. We choose for simplicity $b(y)=w^{-1}M \cosh(2My/w)/\cosh M$ with $M\gg 1$.
An explicit solution of the kinetic equation using this effective mean free path reveals a substantially altered current profile (Fig.~\ref{fig:pathology}). Also the profile of the transverse electric field is affected as shown in Fig.~\ref{fig:crosstohydro}.

\begin{figure}
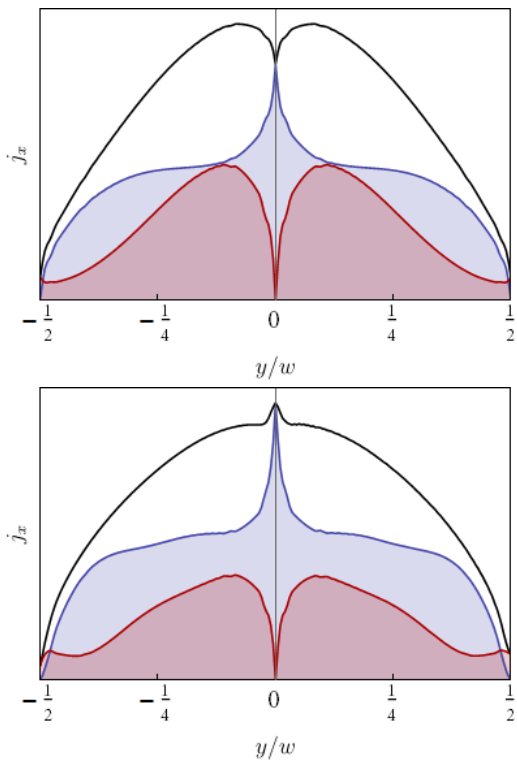

\includegraphics[width=.8\columnwidth]{Fig10a}

\includegraphics[width=.8\columnwidth]{Fig10b}
\caption{Contribution of the various parts of the distribution function to the current profile. Regions strongly affected by the boundary (colored red in Fig.~\ref{fig:characters}) result in the red curve, the bulk part (colored blue in Fig.~\ref{fig:characters}) leads to the blue curve. Both a hard wall boundary condition (a) and a smooth boundary (b) result in the qualitatively similar current contributions, but their sum, (black) sensitively depends on the respective size of both parts.}
\label{fig:bouncenobounce}
\end{figure}

Obviously, a finite size absorptive layer is distinct from a diffusive hard wall. But surprisingly, making the absorptive layer narrower does not necessarily restore the current profile of the hard wall limit, at least to the extent that we were able to test it numerically. To understand the origin of this difference, we emphasize that bouncing from a hard wall is intrinsically an asymmetric process, where all the momentum is lost at the impact istant; On the other hand, the momentum loss when traversing a narrow layer with a decreased mean free path penalizes both the incident and exiting part of the trajectory symmetrically. 
Another important difference is the strength of the diffusive boundary. While a hard wall can at most remove all momentum from the incident particle, a smooth boundary can suppress skipping orbits to any extent and remove them completely. It is this \emph{superdiffusive} behavior which can produce qualitative changes in the profiles of the current and the transverse electric field.
We confirm this picture by dividing the the current contributions into colliding/non-colliding parts, shown in Fig.~\ref{fig:bouncenobounce} in red and blue. While both current parts keep their qualitative shapes, they have essentially the opposite functional form in the middle of the channel. Small changes in the relative sizes will thus result in a very different total current density.

We therefore conclude that a bounce trajectory does not represent the only semiclassical limit of hot electrons colliding with an atomically sized boundary. It can instead be represented by the zero width limit of a finite width absorptive layer with small disorder mean free path. The differentiation only becomes necessary due to the very long mean free path of the electrons away from the boundary and most importantly does neither affect the well-known ohmic nor the hydrodynamic regime. In both cases, bounce trajectories are indistinguishable from a smooth boundary scattering region. We emphasize that the Gurzhi regime is affected as well by these ballistic effects, even though $\ell_{ee}<w$.

For the purposes of the present work, all these considerations have surprisingly little effect, because we are first and foremost interested in distinguishing the various regimes of transport. Indeed, any deviation from the predicted $cosh$-profile of hydrodynamic flow is already a valuable signal and the exact differentiation of bouncy electrons and superdiffusive boundary scattering are only relevant if they lead to a profile which would erroneously resemble the hydrodynamic flow. 
As we explained, while this can happen for the current profile for a smooth boundary condition (Fig.~\ref{fig:pathology}), we did not find a case where this happens for the profile of the transverse electric field (Fig.~\ref{fig:crosstohydro}).
The positive statement we can thus make is that the transverse electric field - but not the current - will invariably, for any tested non-specular boundary condition exhibit an identifiable signal of the onset of correlations beyond local friction. This signal can be measured with a capacitative local probe of the type which they are available now~\cite{Ella2018,Sulpizio2019}.

\section{Conclusions}
\label{sec:five}
We mapped out the properties of the transverse electric field for a Hall geometry with large disorder mean free path. 
Concerning the interaction dominated regime, we identified some important limitations of the standard hydrodynamic approach when combined with the ubiquitous no-slip boundary condition. 

To distinguish the different types of non-ohmic transport, one can use the curvature of the transverse electric field in weak magnetic fields ($w/R_c<2$), which is positive for ballistic flow and negative for hydrodynamic flow.
However, it turned out to be particularly convenient to use a magnetic field strength with $w/R_c=4$, corresponding to cyclotron orbits which completely decouple the skipping orbits of the two channel walls. 
In this case, the profile of the transverse electric field acquires a distinctive triangular shape and depending on the boundary exhibits a pronounced peak or cusp in the center of the channel. This peak becomes smaller with decreasing interaction mean free path while the whole profile changes from triangular to parabolic. The crossover happens when the Gurzhi length $\sqrt{\ell_0\ell}$ becomes smaller than the channel width, below which the established hydrodynamic equations are valid again. 
We emphasize that the Gurzhi regime itself ($\sqrt{\ell_0\ell}\sim w$) is closer to ballistic than to hydrodynamic transport.

The prescribed phenomenology is useful in the experimental study of unconventional transport,
while with the current profile alone it is more delicate to differentiate between ballistic and hydrodynamic flow. 
We also highlighted the unexpectedly quick failure of the hydrodynamic formalism when non-local correlations ranging beyond those associated with hydrodynamic flow are present. Similar observations have been put forward in some holographic models~\cite{Lucas2017}, but a simple experimental signature of this threshold was so far elusive. We expect this phenomenology to be realizable in other almost hydrodynamic systems with parity breaking, like for example in active chiral fluids~\cite{Souslov2019}.
\begin{acknowledgments}
The authors acknowledge support from the Minerva Foundation (T.~H.) and from the Emergent Phenomena in Quantum Systems initiative of the Gordon and Betty Moore Foundation (T.~S.).
\end{acknowledgments}

\setcounter{equation}{0}
\renewcommand{\theequation}{A\arabic{equation}}

\setcounter{section}{0}
\renewcommand{\thesection}{\Alph{section}}

\section{Charging effects}
\label{sec:A}
In the main text we disregarded most effects of charging, in the following we remedy this, but as expected, the overall impact on the transverse electric field is negligible already at moderately low carrier densities.
The discussion is still simplified in the sense that the 2D channel is assumed not to be compensated by a backgate, but is otherwise general. 
Denoting the coordinate perpendicular to the plane of current flow with $z$, Gauss's law for a planar charge density $\rho(y)$ reads
\begin{align}
    \partial_y E_y(y,z)+\partial_z E_z(y,z)=4\pi \rho(y) \delta(z).
\end{align}
This has no easy solution in the plane as $E_z(y,z)$ jumps to its negative at $z=0$. We instead take the direct route and integrate
\begin{align}
    E_y(y,z)&=-\partial_y\int\!\dd^3r' \frac{\rho(y')\delta(z)}{|\bm{r}-\bm{r'}|^p}\\
    E_y(y,0)&=\int\!\dd y'\frac{2\rho(y')}{y-y'},
    \label{eq:coulomb}
\end{align}
where we used $p=1$ which is true without backgate.
As outlined before, the determination of the electrostatic situation is not completely decoupled from the current profile. However, it is often a good working assumption that the correct density profile $h_0$ does not change the current profile $h_1^e$ very much. We are now in the position to quantify this statement.
The density profile is determined by the consistency requirement
\begin{align}
    \frac{h_1^e(y)}{R_c}+&\tfrac{1}{2}\partial_y h_2^e(y)
    \notag\\
    &=\partial_yh_0(y)+\frac{\kappa_{TF}}{2\pi}\int\!\dd y'\frac{ h_0(y')+h_0^{add}(y')}{y-y'}
    \label{eq:gausslaw}
\end{align}
where we have introduced the Thomas-Fermi screening wavevector $\kappa_{TF}=2e^2m/\varepsilon\hbar^2$ and accounted for a dielectric constant $\varepsilon$ and some additional charges $h_0^{add}$ residing outside of the channel. Using for the left hand side the solution of the kinetic equation which we determined previously, this equation can be solved for $h_0$. 
To this end, we extend Eq.~\eqref{eq:gausslaw} over the entire $y$-axis, combining $h_0+h_0^{add}\to h_0$, and include a chemical potential shift also for charges outside the channel. Electric fields outside the channel are accounted for as $E_y^{ext}$. Now, the Coulomb integral assumes the form of a Hilbert transform $\mathcal{H}$, which can be inverted,
\begin{align}
   h_1^e(y)/R_c+\tfrac{1}{2}\partial_y h_2^e+E_y^{ext}
    &=\partial_yh_0+\frac{\kappa_{TF}}{2}\mathcal{H}(h_0)
    \label{eq:hilberted}\\
   \mathcal{H}(h_1^e/R_c+\tfrac{1}{2}\partial_y h_2^e+E^{ext})
    &=\partial_y\mathcal{H}(h_0)-\tfrac{\kappa_{TF}}{2}h_0
    \label{eq:inversehilbert}
\end{align}
where we used the properties of the Hilbert transform to pull out the derivative~\cite{Pandey1996}. Inserting Eq~\eqref{eq:hilberted} for $\mathcal{H}(h_0(y))$ leads to a differential equation in $h_0$,
\begin{align}
   \tfrac{2}{\kappa_{TF}}&(\partial_yh_1^e/R_c
   +\tfrac{1}{2}\partial_y^2 h_2^e+\partial_yE_y^{ext}
   -\partial_y^2h_0)
   -\tfrac{\kappa_{TF}}{2}h_0
   \notag\\
   &=\mathcal{H}(h_1^e/R_c+\tfrac{1}{2}\partial_y h_2^e+E^{ext})
\end{align}
Using a gradient expansion, to zeroth order this reduces to
\begin{align}
   \mathcal{H}(h_1^e/R_c+E^{ext})
    &=-\tfrac{\kappa_{TF}}{2}h_0.
\end{align}
If stress is not negligible, one should instead expand to first order. We also emphasize that $\kappa_{TF} w\gg1$, which is true even for moderate number densities, is not enough by itself as all terms in Eq.~\eqref{eq:inversehilbert} are of comparable size.
We look at two representative cases, one without charges in the proximity, and secondly the case of full compensation at the boundaries, where the electric field is zero for $|y|>w/2$. 
Without additional charges outside the channel, the electric field can potentially become very large at the boundary. Taylor expanding $h_0$ in Eq.~\eqref{eq:coulomb} around one of the boundaries, it becomes obvious that
a nonzero but finite charge on the boundary implies a divergence of the electric field there. This is undesirable, therefore we impose $h_0(\pm w/2)=0$. Without any outside charges, the electric field on the boundary is definitely negative. Such a solution is perfectly possible, but unless the density of states is pretty high, requires solving the kinetic equation anew. The reason is the correction to the electric field at the boundary where the current is small anyway, quickly goes beyond the negligible backreaction that is typically assumed for charging. 

\begin{figure}
    \centering
    \includegraphics[width=.9\columnwidth]{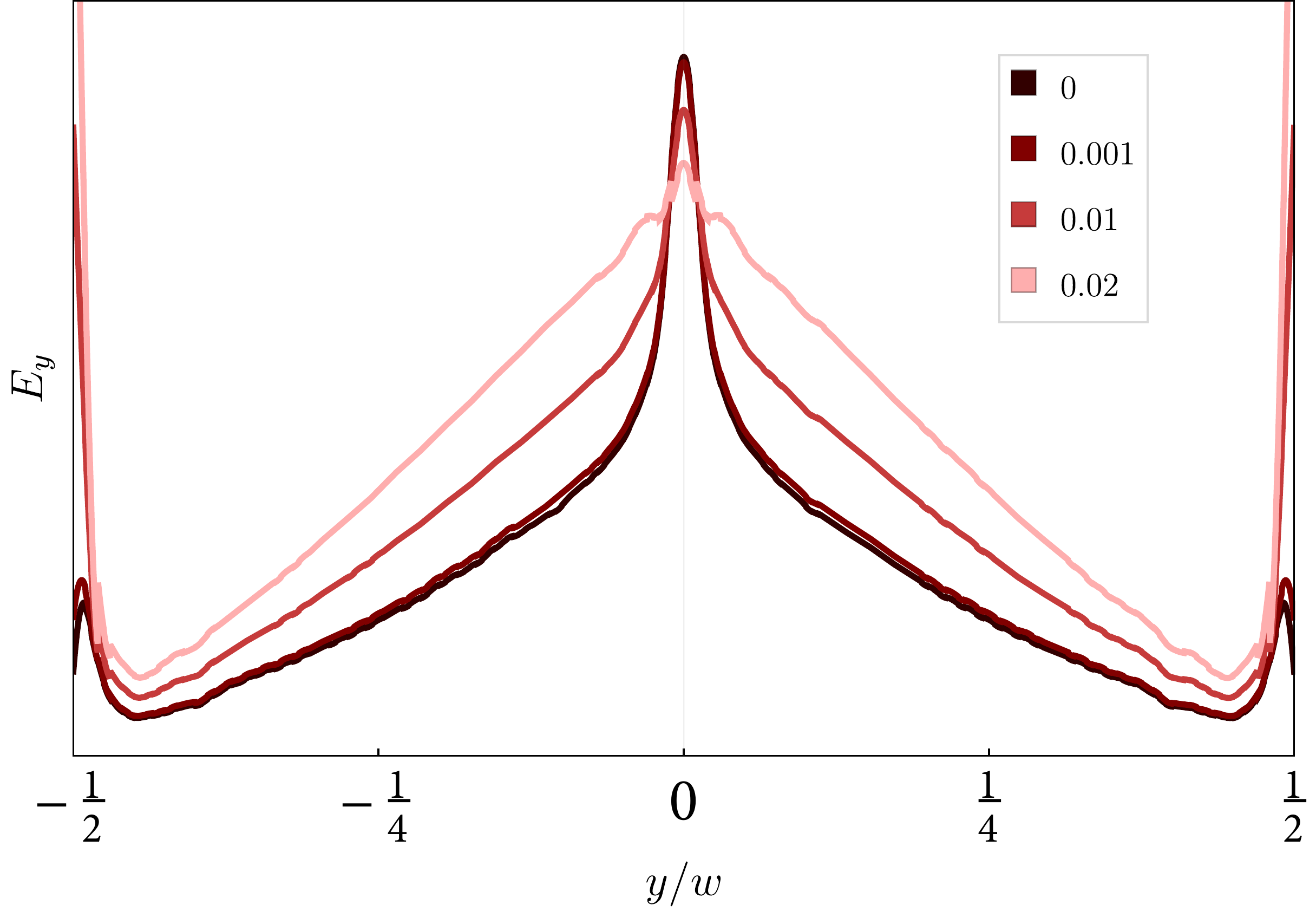}
    
    \includegraphics[width=.9\columnwidth]{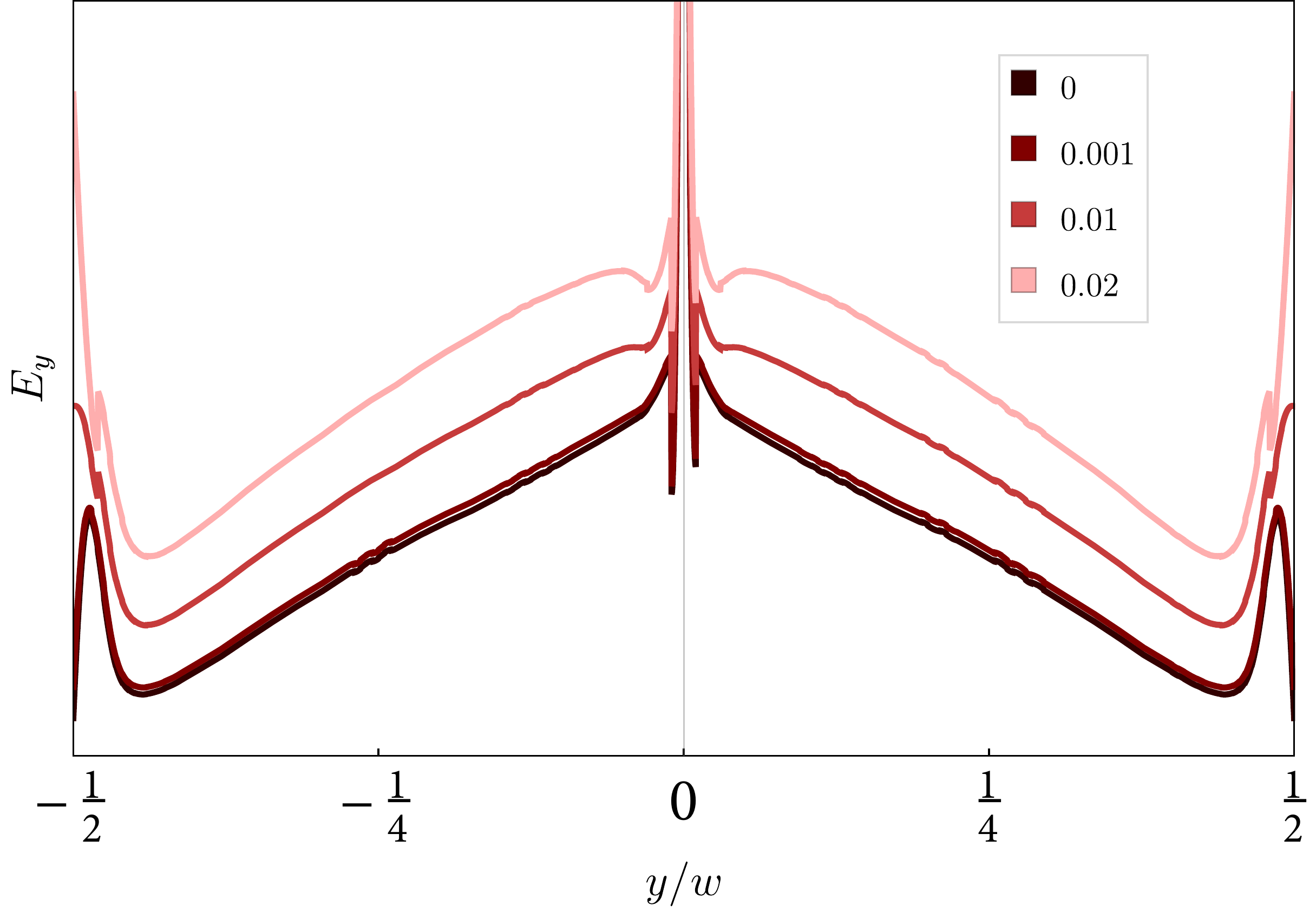}
    \caption{Ballistic transport: Sensitivity of the transverse electric field on boundary conditions and the electron density at $w/R_c=4$. 
    The left panel (a) shows $E_y$ for reflectively $r_h=0$ with characteristic peaks at the center and boundaries of the channel at high densities, which get attenuated for very low carrier densities. In (b) we consider instead a narrow absorptive boundary layer of width $w/M=w/200$. In this case, the central peak becomes very narrow, almost insensitive to a decreased screening wavelength. On the other had the peak at the wall is not strongly affected. On the right panel,  the magnetic field was slightly increased by $2/M=1\%$ to account for the smaller effective channel width. The other parameters are $l_0=3w$, $l_{ee}=7w$.}
    \label{fig:Eywithkappa}
\end{figure}

In the second case without external fields, using the mode expansion and to first order in gradients we obtain for the electric field
\begin{align}
    E_y&=\tfrac{1}{R_c}h_1^e+\tfrac{1}{2}\partial_yh_2^e
    +\tfrac{2}{\kappa_{TF} R_c}\partial_y\mathcal{H}(h_1^e).
    \label{eq:completeEy}
\end{align}
Naturally, the correction due to charging will become more important at lower carrier densities.
We use this result to reconstruct the electric field from the  solution of the kinetic equation for some representative densities in Fig.~\ref{fig:Eywithkappa}. 
Importantly, using a diffusive boundary condition, at finite magnetic fields the correction due to the density gradient can become sizable and even change the spatial dependence of $E_y$. Nonetheless, it is generally suppressed very fast with increasing carrier density. We emphasize that the the profile of $E_y$ can be used to differentiate between hydrodynamic and ballistic transport in exactly the same manner as detailed in the main text even in the presence of electrostatic effects.

\section{Curvature of the transverse field in weak magnetic fields}
\label{sec:B}
In the following, we present the calculation of the normalized curvature.
The parameters are $l_{ee}\to\infty$, $l_{0}\to\infty$ and $R_c\gg w$.
In this limit, numerically the transverse electric field is found to be mostly independent of $y$. Therefore, we attempt to solve the Boltzmann equation with the simple ansatz $E_y(y)=E_y^{(0)}$. For fully diffusive walls, $r_h=0$, the distribution function can be divided into two symmetric smooth parts ${\tilde h}(y,\theta)$  defined for angles between $\theta_i(y)=\arccos(1-(w/2+y)/R_c)$ and $\theta_f(y)=2\pi-\arccos((w/2-y)/R_c-1)$. The full distribution function is constructed by the combination
\begin{align}
    h(y,\theta)
    &=
    {\tilde h}(+y,\theta)+{\tilde h}(-y,\pi-\theta).
\end{align}
The solution to the kinetic equation for $\ell_{ee}\to\infty$ and $w/R_c<2$ is given by
\begin{widetext}
\begin{align}
    {\tilde h}(y,\theta)
    &=\frac{l_0}{1+l_0^2/R_c^2}
    \Big(E_x
    \left(
    \cos\theta-A
    \left(
    u+\sqrt{1-u^2} l_0/R_c
    \right)
    +\sin\theta l_0/R_c
    \right)
    \notag\\&
    -E_y^{(0)}
    \left(
    A
    \left(
    u l_0/R_c-\sqrt{1-u^2}
    \right)
    -\cos\theta l_0/R_c+\sin\theta
    \right)\Big)
\end{align}
where $u=(w/2+y)/R_c+\cos\theta$ and $A=\exp((\arccos u-\theta)R_c/l_0)$. This solution is defined for angles between $\theta_i(y)=\arccos(1-(w/2+y)/R_c)$ and $\theta_f(y)=2\pi-\arccos((w/2-y)/R_c-1)$.

Expanding for $l_0\to\infty$, this results in Eq.~\eqref{eq:infinitel0} of the main text
\begin{align}
    {\tilde h}(y,\theta)&=
    E_x R_c\left(
    \sin\theta-\sqrt{1-\left(\frac{\tfrac{w}{2}+y}{R_c}+\cos\theta\right)^2}
    \right)
    -E_y^{(0)}(w/2+y).
\end{align}
We calculated $h_0$, $h_1^e$ and $h_2^e$ from this expression.
The density is calculated as 
\begin{align}
    \frac{2\pi h_0}{E_x R_c}&=
    \int_{\theta_i(y)}^{\theta_f(y)}\!\!
    \dd\theta{\tilde h}(y,\theta)
    -(y\to -y)
    \\
    &=-1+\alpha_2-(1-\alpha_1)
    -\int_{1-\alpha_1}^{-1}\!\!\dd\alpha
    \frac{\sqrt{1-\left(\frac{\tfrac{w}{2}+y}{R_c}+\alpha\right)^2}}
    {\sqrt{1-\alpha^2}}
    -\int_{-1}^{-1+\alpha_2}\!\!\dd\alpha
    \frac{\sqrt{1-\left(\frac{\tfrac{w}{2}+y}{R_c}+\alpha\right)^2}}
    {\sqrt{1-\alpha^2}}-(y\to -y)
    \shortintertext{where it is $1-\alpha_1=\cos\theta_i(y)=1-(w/2+y)/R_c$ and $-1+\alpha_2=\cos\theta_f(y)=-1+(w/2-y)/R_c$. For $R_c\to\infty$, we can further split up the integration regions and expand for $\alpha\approx \pm 1$,}
    &\approx
    \frac{w}{R_c}-2
    +\int_0^1\!\!\dd\alpha
    \frac{\sqrt{1-\alpha+\alpha_1}}{\sqrt{1-\alpha}}
    -\int_{1-\alpha_2}^1\!\!\dd\alpha
    \frac{\sqrt{1-\alpha+\alpha_1}}{\sqrt{1-\alpha}}
    +\int_{0}^{1-\alpha_1}\!\!\dd\alpha
    \frac{\sqrt{1-\alpha-\alpha_1}}{\sqrt{1-\alpha}}
    -(y\to -y)
    \\
    &=\left(
    \sqrt{1+\alpha_1}-\alpha_1\log\frac{\sqrt{\alpha_1}}{1+\sqrt{1+\alpha_1}}
    \right)
    +\left(
    \alpha_1\log\frac{\sqrt{\alpha_1}}{\sqrt{\alpha_2}+\sqrt{\alpha_1+\alpha_2}}-\sqrt{\alpha_2(\alpha_1+\alpha_2)}
    \right)
    \notag\\&\quad
    +\left(
    \sqrt{1-\alpha_1}+\alpha_1\log\frac{\sqrt{\alpha_1}}{1+\sqrt{1-\alpha_1}}
    \right)
    -(y\to -y)
    \\
     h_0&\approx
    \frac{E_x}{2\pi}\left(
    (w/2+y)\log\frac{\sqrt{w/2+y}}{\sqrt{w/2-y}+\sqrt{w/2}}
    -\sqrt{(w/2-y)w/2}-(y\to -y)
    \right)+\mathcal{O}(R_c^{-1})
    \\
    &\approx
    \frac{E_x}{\pi}\left((\sqrt{2}-\log(1+\sqrt{2}))y+\frac{\sqrt{2}}{3w^2}y^3
    \right)+\mathcal{O}(y^5)
\end{align}
The same procedure leads to the viscous contribution
\begin{align}
    \frac{\partial_y}{2} h_2^e
    &=
    \frac{\partial_y}{2}
    \int_{\theta_i(y)}^{\theta_f(y)}\!\!
    \frac{\dd\theta}{\pi}\cos(2\theta){\tilde h}(y,\theta)
    +(y\to -y)
    \notag\\
    &\approx
    \frac{E_x}{\pi}\left(\sqrt{2}-\log(1+\sqrt{2})+\frac{\sqrt{2}}{w^2}y^2
    \right)+\mathcal{O}(y^4).
\end{align}
And finally, the current is approximately
\begin{align}
    h_1^e&=
    \frac{E_x w}{\pi}\left(
    \log\frac{8R_c}{w}+\log(1+\sqrt{2})-\sqrt{2}-(2+\sqrt{2})\frac{y^2}{w^2}
    \right)+\mathcal{O}(y^4).
\end{align}
\end{widetext}
The electric field is defined as $E_y=h_1^e(y)/R_c+\tfrac{1}{2}\partial_y h_2^e$. For small magnetic fields, the viscous term $\partial_y h_2^e$ of order $E_x$ is the dominant contribution to $E_y$ compared to the ohmic term $h_1^e(y)/R_c$ which is smaller by a factor $R_c$. Inserting all numbers yields explicitly
\begin{align}
    \frac{E_y^{(2)}}{2E_y^{(0)}}
    &=\frac{\sqrt{2}}{\sqrt{2}-\log(1+\sqrt{2})}\frac{1}{w^2}\approx
    \frac{2.65}{w^2},
\end{align}
which is the result quoted in the main text.


\begin{thebibliography}{52}%
\makeatletter
\providecommand \@ifxundefined [1]{%
 \@ifx{#1\undefined}
}%
\providecommand \@ifnum [1]{%
 \ifnum #1\expandafter \@firstoftwo
 \else \expandafter \@secondoftwo
 \fi
}%
\providecommand \@ifx [1]{%
 \ifx #1\expandafter \@firstoftwo
 \else \expandafter \@secondoftwo
 \fi
}%
\providecommand \natexlab [1]{#1}%
\providecommand \enquote  [1]{``#1''}%
\providecommand \bibnamefont  [1]{#1}%
\providecommand \bibfnamefont [1]{#1}%
\providecommand \citenamefont [1]{#1}%
\providecommand \href@noop [0]{\@secondoftwo}%
\providecommand \href [0]{\begingroup \@sanitize@url \@href}%
\providecommand \@href[1]{\@@startlink{#1}\@@href}%
\providecommand \@@href[1]{\endgroup#1\@@endlink}%
\providecommand \@sanitize@url [0]{\catcode `\\12\catcode `\$12\catcode
  `\&12\catcode `\#12\catcode `\^12\catcode `\_12\catcode `\%12\relax}%
\providecommand \@@startlink[1]{}%
\providecommand \@@endlink[0]{}%
\providecommand \url  [0]{\begingroup\@sanitize@url \@url }%
\providecommand \@url [1]{\endgroup\@href {#1}{\urlprefix }}%
\providecommand \urlprefix  [0]{URL }%
\providecommand \Eprint [0]{\href }%
\providecommand \doibase [0]{http://dx.doi.org/}%
\providecommand \selectlanguage [0]{\@gobble}%
\providecommand \bibinfo  [0]{\@secondoftwo}%
\providecommand \bibfield  [0]{\@secondoftwo}%
\providecommand \translation [1]{[#1]}%
\providecommand \BibitemOpen [0]{}%
\providecommand \bibitemStop [0]{}%
\providecommand \bibitemNoStop [0]{.\EOS\space}%
\providecommand \EOS [0]{\spacefactor3000\relax}%
\providecommand \BibitemShut  [1]{\csname bibitem#1\endcsname}%
\let\auto@bib@innerbib\@empty
\bibitem [{\citenamefont {{de Jong}}\ and\ \citenamefont
  {{Molenkamp}}(1995)}]{deJong1995}%
  \BibitemOpen
  \bibfield  {author} {\bibinfo {author} {\bibfnamefont {M.~J.~M.}\
  \bibnamefont {{de Jong}}}\ and\ \bibinfo {author} {\bibfnamefont {L.~W.}\
  \bibnamefont {{Molenkamp}}},\ }\href {\doibase 10.1103/PhysRevB.51.13389}
  {\bibfield  {journal} {\bibinfo  {journal} {Phys. Rev. B}\ }\textbf {\bibinfo
  {volume} {51}},\ \bibinfo {pages} {13389} (\bibinfo {year}
  {1995})}\BibitemShut {NoStop}%
\bibitem [{\citenamefont {{M{\"u}ller}}\ and\ \citenamefont
  {{Sachdev}}(2008)}]{Mueller2008}%
  \BibitemOpen
  \bibfield  {author} {\bibinfo {author} {\bibfnamefont {M.}~\bibnamefont
  {{M{\"u}ller}}}\ and\ \bibinfo {author} {\bibfnamefont {S.}~\bibnamefont
  {{Sachdev}}},\ }\href {\doibase 10.1103/PhysRevB.78.115419} {\bibfield
  {journal} {\bibinfo  {journal} {Phys. Rev. B}\ }\textbf {\bibinfo {volume}
  {78}},\ \bibinfo {pages} {115419} (\bibinfo {year} {2008})}\BibitemShut
  {NoStop}%
\bibitem [{\citenamefont {{Fritz}}\ \emph {et~al.}(2008)\citenamefont
  {{Fritz}}, \citenamefont {{Schmalian}}, \citenamefont {{M{\"u}ller}},\ and\
  \citenamefont {{Sachdev}}}]{Fritz2008}%
  \BibitemOpen
  \bibfield  {author} {\bibinfo {author} {\bibfnamefont {L.}~\bibnamefont
  {{Fritz}}}, \bibinfo {author} {\bibfnamefont {J.}~\bibnamefont
  {{Schmalian}}}, \bibinfo {author} {\bibfnamefont {M.}~\bibnamefont
  {{M{\"u}ller}}}, \ and\ \bibinfo {author} {\bibfnamefont {S.}~\bibnamefont
  {{Sachdev}}},\ }\href {\doibase 10.1103/PhysRevB.78.085416} {\bibfield
  {journal} {\bibinfo  {journal} {Phys. Rev. B}\ }\textbf {\bibinfo {volume}
  {78}},\ \bibinfo {pages} {085416} (\bibinfo {year} {2008})}\BibitemShut
  {NoStop}%
\bibitem [{\citenamefont {{M{\"u}ller}}\ \emph {et~al.}(2009)\citenamefont
  {{M{\"u}ller}}, \citenamefont {{Schmalian}},\ and\ \citenamefont
  {{Fritz}}}]{Mueller2009}%
  \BibitemOpen
  \bibfield  {author} {\bibinfo {author} {\bibfnamefont {M.}~\bibnamefont
  {{M{\"u}ller}}}, \bibinfo {author} {\bibfnamefont {J.}~\bibnamefont
  {{Schmalian}}}, \ and\ \bibinfo {author} {\bibfnamefont {L.}~\bibnamefont
  {{Fritz}}},\ }\href {\doibase 10.1103/PhysRevLett.103.025301} {\bibfield
  {journal} {\bibinfo  {journal} {Phys. Rev. Lett.}\ }\textbf {\bibinfo
  {volume} {103}},\ \bibinfo {pages} {025301} (\bibinfo {year}
  {2009})}\BibitemShut {NoStop}%
\bibitem [{\citenamefont {{Bistritzer}}\ and\ \citenamefont
  {{MacDonald}}(2009)}]{Bistritzer2009}%
  \BibitemOpen
  \bibfield  {author} {\bibinfo {author} {\bibfnamefont {R.}~\bibnamefont
  {{Bistritzer}}}\ and\ \bibinfo {author} {\bibfnamefont {A.~H.}\ \bibnamefont
  {{MacDonald}}},\ }\href {\doibase 10.1103/PhysRevB.80.085109} {\bibfield
  {journal} {\bibinfo  {journal} {Phys. Rev. B}\ }\textbf {\bibinfo {volume}
  {80}},\ \bibinfo {pages} {085109} (\bibinfo {year} {2009})}\BibitemShut
  {NoStop}%
\bibitem [{\citenamefont {{Mendoza}}\ \emph {et~al.}(2011)\citenamefont
  {{Mendoza}}, \citenamefont {{Herrmann}},\ and\ \citenamefont
  {{Succi}}}]{Mendoza2011}%
  \BibitemOpen
  \bibfield  {author} {\bibinfo {author} {\bibfnamefont {M.}~\bibnamefont
  {{Mendoza}}}, \bibinfo {author} {\bibfnamefont {H.~J.}\ \bibnamefont
  {{Herrmann}}}, \ and\ \bibinfo {author} {\bibfnamefont {S.}~\bibnamefont
  {{Succi}}},\ }\href {\doibase 10.1103/PhysRevLett.106.156601} {\bibfield
  {journal} {\bibinfo  {journal} {Phys. Rev. Lett.}\ }\textbf {\bibinfo
  {volume} {106}},\ \bibinfo {pages} {156601} (\bibinfo {year}
  {2011})}\BibitemShut {NoStop}%
\bibitem [{\citenamefont {{Mendoza}}\ \emph {et~al.}(2013)\citenamefont
  {{Mendoza}}, \citenamefont {{Herrmann}},\ and\ \citenamefont
  {{Succi}}}]{Mendoza2013}%
  \BibitemOpen
  \bibfield  {author} {\bibinfo {author} {\bibfnamefont {M.}~\bibnamefont
  {{Mendoza}}}, \bibinfo {author} {\bibfnamefont {H.~J.}\ \bibnamefont
  {{Herrmann}}}, \ and\ \bibinfo {author} {\bibfnamefont {S.}~\bibnamefont
  {{Succi}}},\ }\href {\doibase 10.1038/srep01052} {\bibfield  {journal}
  {\bibinfo  {journal} {Scientific Reports}\ }\textbf {\bibinfo {volume} {3}},\
  \bibinfo {pages} {1052} (\bibinfo {year} {2013})}\BibitemShut {NoStop}%
\bibitem [{\citenamefont {{Narozhny}}\ \emph {et~al.}(2015)\citenamefont
  {{Narozhny}}, \citenamefont {{Gornyi}}, \citenamefont {{Titov}},
  \citenamefont {{Sch{\"u}tt}},\ and\ \citenamefont {{Mirlin}}}]{Narozhny2015}%
  \BibitemOpen
  \bibfield  {author} {\bibinfo {author} {\bibfnamefont {B.~N.}\ \bibnamefont
  {{Narozhny}}}, \bibinfo {author} {\bibfnamefont {I.~V.}\ \bibnamefont
  {{Gornyi}}}, \bibinfo {author} {\bibfnamefont {M.}~\bibnamefont {{Titov}}},
  \bibinfo {author} {\bibfnamefont {M.}~\bibnamefont {{Sch{\"u}tt}}}, \ and\
  \bibinfo {author} {\bibfnamefont {A.~D.}\ \bibnamefont {{Mirlin}}},\ }\href
  {\doibase 10.1103/PhysRevB.91.035414} {\bibfield  {journal} {\bibinfo
  {journal} {Phys. Rev. B}\ }\textbf {\bibinfo {volume} {91}},\ \bibinfo
  {pages} {035414} (\bibinfo {year} {2015})}\BibitemShut {NoStop}%
\bibitem [{\citenamefont {{Briskot}}\ \emph {et~al.}(2015)\citenamefont
  {{Briskot}}, \citenamefont {{Sch{\"u}tt}}, \citenamefont {{Gornyi}},
  \citenamefont {{Titov}}, \citenamefont {{Narozhny}},\ and\ \citenamefont
  {{Mirlin}}}]{Briskot2015}%
  \BibitemOpen
  \bibfield  {author} {\bibinfo {author} {\bibfnamefont {U.}~\bibnamefont
  {{Briskot}}}, \bibinfo {author} {\bibfnamefont {M.}~\bibnamefont
  {{Sch{\"u}tt}}}, \bibinfo {author} {\bibfnamefont {I.~V.}\ \bibnamefont
  {{Gornyi}}}, \bibinfo {author} {\bibfnamefont {M.}~\bibnamefont {{Titov}}},
  \bibinfo {author} {\bibfnamefont {B.~N.}\ \bibnamefont {{Narozhny}}}, \ and\
  \bibinfo {author} {\bibfnamefont {A.~D.}\ \bibnamefont {{Mirlin}}},\ }\href
  {\doibase 10.1103/PhysRevB.92.115426} {\bibfield  {journal} {\bibinfo
  {journal} {Phys. Rev. B}\ }\textbf {\bibinfo {volume} {92}},\ \bibinfo
  {pages} {115426} (\bibinfo {year} {2015})}\BibitemShut {NoStop}%
\bibitem [{\citenamefont {{Lucas}}\ \emph {et~al.}(2016)\citenamefont
  {{Lucas}}, \citenamefont {{Crossno}}, \citenamefont {{Fong}}, \citenamefont
  {{Kim}},\ and\ \citenamefont {{Sachdev}}}]{Lucas2016}%
  \BibitemOpen
  \bibfield  {author} {\bibinfo {author} {\bibfnamefont {A.}~\bibnamefont
  {{Lucas}}}, \bibinfo {author} {\bibfnamefont {J.}~\bibnamefont {{Crossno}}},
  \bibinfo {author} {\bibfnamefont {K.~C.}\ \bibnamefont {{Fong}}}, \bibinfo
  {author} {\bibfnamefont {P.}~\bibnamefont {{Kim}}}, \ and\ \bibinfo {author}
  {\bibfnamefont {S.}~\bibnamefont {{Sachdev}}},\ }\href {\doibase
  10.1103/PhysRevB.93.075426} {\bibfield  {journal} {\bibinfo  {journal} {Phys.
  Rev. B}\ }\textbf {\bibinfo {volume} {93}},\ \bibinfo {pages} {075426}
  (\bibinfo {year} {2016})}\BibitemShut {NoStop}%
\bibitem [{\citenamefont {{Armitage}}\ \emph {et~al.}(2018)\citenamefont
  {{Armitage}}, \citenamefont {{Mele}},\ and\ \citenamefont
  {{Vishwanath}}}]{Armitage2018}%
  \BibitemOpen
  \bibfield  {author} {\bibinfo {author} {\bibfnamefont {N.~P.}\ \bibnamefont
  {{Armitage}}}, \bibinfo {author} {\bibfnamefont {E.~J.}\ \bibnamefont
  {{Mele}}}, \ and\ \bibinfo {author} {\bibfnamefont {A.}~\bibnamefont
  {{Vishwanath}}},\ }\href {\doibase 10.1103/RevModPhys.90.015001} {\bibfield
  {journal} {\bibinfo  {journal} {Rev. Mod. Phys.}\ }\textbf {\bibinfo {volume}
  {90}},\ \bibinfo {pages} {015001} (\bibinfo {year} {2018})}\BibitemShut
  {NoStop}%
\bibitem [{\citenamefont {{Gurzhi}}(1968)}]{Gurzhi1968}%
  \BibitemOpen
  \bibfield  {author} {\bibinfo {author} {\bibfnamefont {R.~N.}\ \bibnamefont
  {{Gurzhi}}},\ }\href {\doibase 10.1070/PU1968v011n02ABEH003815} {\bibfield
  {journal} {\bibinfo  {journal} {Soviet Physics Uspekhi}\ }\textbf {\bibinfo
  {volume} {11}},\ \bibinfo {pages} {255} (\bibinfo {year} {1968})}\BibitemShut
  {NoStop}%
\bibitem [{\citenamefont {Beenakker}\ and\ \citenamefont {van
  Houten}(1991)}]{Beenakker1991}%
  \BibitemOpen
  \bibfield  {author} {\bibinfo {author} {\bibfnamefont {C.}~\bibnamefont
  {Beenakker}}\ and\ \bibinfo {author} {\bibfnamefont {H.}~\bibnamefont {van
  Houten}},\ }in\ \href {\doibase
  https://doi.org/10.1016/S0081-1947(08)60091-0} {\emph {\bibinfo {booktitle}
  {Semiconductor Heterostructures and Nanostructures}}},\ \bibinfo {series}
  {Solid State Physics}, Vol.~\bibinfo {volume} {44},\ \bibinfo {editor}
  {edited by\ \bibinfo {editor} {\bibfnamefont {H.}~\bibnamefont {Ehrenreich}}\
  and\ \bibinfo {editor} {\bibfnamefont {D.}~\bibnamefont {Turnbull}}}\
  (\bibinfo  {publisher} {Academic Press},\ \bibinfo {year} {1991})\ pp.\
  \bibinfo {pages} {1 -- 228}\BibitemShut {NoStop}%
\bibitem [{\citenamefont {{Bandurin}}\ \emph {et~al.}(2016)\citenamefont
  {{Bandurin}}, \citenamefont {{Torre}}, \citenamefont {{Kumar}}, \citenamefont
  {{Ben Shalom}}, \citenamefont {{Tomadin}}, \citenamefont {{Principi}},
  \citenamefont {{Auton}}, \citenamefont {{Khestanova}}, \citenamefont
  {{Novoselov}}, \citenamefont {{Grigorieva}}, \citenamefont {{Ponomarenko}},
  \citenamefont {{Geim}},\ and\ \citenamefont {{Polini}}}]{Bandurin2016}%
  \BibitemOpen
  \bibfield  {author} {\bibinfo {author} {\bibfnamefont {D.~A.}\ \bibnamefont
  {{Bandurin}}}, \bibinfo {author} {\bibfnamefont {I.}~\bibnamefont {{Torre}}},
  \bibinfo {author} {\bibfnamefont {R.~K.}\ \bibnamefont {{Kumar}}}, \bibinfo
  {author} {\bibfnamefont {M.}~\bibnamefont {{Ben Shalom}}}, \bibinfo {author}
  {\bibfnamefont {A.}~\bibnamefont {{Tomadin}}}, \bibinfo {author}
  {\bibfnamefont {A.}~\bibnamefont {{Principi}}}, \bibinfo {author}
  {\bibfnamefont {G.~H.}\ \bibnamefont {{Auton}}}, \bibinfo {author}
  {\bibfnamefont {E.}~\bibnamefont {{Khestanova}}}, \bibinfo {author}
  {\bibfnamefont {K.~S.}\ \bibnamefont {{Novoselov}}}, \bibinfo {author}
  {\bibfnamefont {I.~V.}\ \bibnamefont {{Grigorieva}}}, \bibinfo {author}
  {\bibfnamefont {L.~A.}\ \bibnamefont {{Ponomarenko}}}, \bibinfo {author}
  {\bibfnamefont {A.~K.}\ \bibnamefont {{Geim}}}, \ and\ \bibinfo {author}
  {\bibfnamefont {M.}~\bibnamefont {{Polini}}},\ }\href {\doibase
  10.1126/science.aad0201} {\bibfield  {journal} {\bibinfo  {journal}
  {Science}\ }\textbf {\bibinfo {volume} {351}},\ \bibinfo {pages} {1055}
  (\bibinfo {year} {2016})}\BibitemShut {NoStop}%
\bibitem [{\citenamefont {{Krishna Kumar}}\ \emph {et~al.}(2017)\citenamefont
  {{Krishna Kumar}}, \citenamefont {{Bandurin}}, \citenamefont {{Pellegrino}},
  \citenamefont {{Cao}}, \citenamefont {{Principi}}, \citenamefont {{Guo}},
  \citenamefont {{Auton}}, \citenamefont {{Ben Shalom}}, \citenamefont
  {{Ponomarenko}}, \citenamefont {{Falkovich}}, \citenamefont {{Watanabe}},
  \citenamefont {{Taniguchi}}, \citenamefont {{Grigorieva}}, \citenamefont
  {{Levitov}}, \citenamefont {{Polini}},\ and\ \citenamefont
  {{Geim}}}]{KrishnaKumar2017}%
  \BibitemOpen
  \bibfield  {author} {\bibinfo {author} {\bibfnamefont {R.}~\bibnamefont
  {{Krishna Kumar}}}, \bibinfo {author} {\bibfnamefont {D.~A.}\ \bibnamefont
  {{Bandurin}}}, \bibinfo {author} {\bibfnamefont {F.~M.~D.}\ \bibnamefont
  {{Pellegrino}}}, \bibinfo {author} {\bibfnamefont {Y.}~\bibnamefont {{Cao}}},
  \bibinfo {author} {\bibfnamefont {A.}~\bibnamefont {{Principi}}}, \bibinfo
  {author} {\bibfnamefont {H.}~\bibnamefont {{Guo}}}, \bibinfo {author}
  {\bibfnamefont {G.~H.}\ \bibnamefont {{Auton}}}, \bibinfo {author}
  {\bibfnamefont {M.}~\bibnamefont {{Ben Shalom}}}, \bibinfo {author}
  {\bibfnamefont {L.~A.}\ \bibnamefont {{Ponomarenko}}}, \bibinfo {author}
  {\bibfnamefont {G.}~\bibnamefont {{Falkovich}}}, \bibinfo {author}
  {\bibfnamefont {K.}~\bibnamefont {{Watanabe}}}, \bibinfo {author}
  {\bibfnamefont {T.}~\bibnamefont {{Taniguchi}}}, \bibinfo {author}
  {\bibfnamefont {I.~V.}\ \bibnamefont {{Grigorieva}}}, \bibinfo {author}
  {\bibfnamefont {L.~S.}\ \bibnamefont {{Levitov}}}, \bibinfo {author}
  {\bibfnamefont {M.}~\bibnamefont {{Polini}}}, \ and\ \bibinfo {author}
  {\bibfnamefont {A.~K.}\ \bibnamefont {{Geim}}},\ }\href {\doibase
  10.1038/nphys4240} {\bibfield  {journal} {\bibinfo  {journal} {Nat. Phys.}\
  }\textbf {\bibinfo {volume} {13}},\ \bibinfo {pages} {1182} (\bibinfo {year}
  {2017})}\BibitemShut {NoStop}%
\bibitem [{\citenamefont {{Moll}}\ \emph {et~al.}(2016)\citenamefont {{Moll}},
  \citenamefont {{Kushwaha}}, \citenamefont {{Nandi}}, \citenamefont
  {{Schmidt}},\ and\ \citenamefont {{Mackenzie}}}]{Moll2016}%
  \BibitemOpen
  \bibfield  {author} {\bibinfo {author} {\bibfnamefont {P.~J.~W.}\
  \bibnamefont {{Moll}}}, \bibinfo {author} {\bibfnamefont {P.}~\bibnamefont
  {{Kushwaha}}}, \bibinfo {author} {\bibfnamefont {N.}~\bibnamefont {{Nandi}}},
  \bibinfo {author} {\bibfnamefont {B.}~\bibnamefont {{Schmidt}}}, \ and\
  \bibinfo {author} {\bibfnamefont {A.~P.}\ \bibnamefont {{Mackenzie}}},\
  }\href {\doibase 10.1126/science.aac8385} {\bibfield  {journal} {\bibinfo
  {journal} {Science}\ }\textbf {\bibinfo {volume} {351}},\ \bibinfo {pages}
  {1061} (\bibinfo {year} {2016})}\BibitemShut {NoStop}%
\bibitem [{\citenamefont {{Gooth}}\ \emph {et~al.}(2018)\citenamefont
  {{Gooth}}, \citenamefont {{Menges}}, \citenamefont {{Kumar}}, \citenamefont
  {{S{\"u}{\ss}}}, \citenamefont {{Shekhar}}, \citenamefont {{Sun}},
  \citenamefont {{Drechsler}}, \citenamefont {{Zierold}}, \citenamefont
  {{Felser}},\ and\ \citenamefont {{Gotsmann}}}]{Gooth2017}%
  \BibitemOpen
  \bibfield  {author} {\bibinfo {author} {\bibfnamefont {J.}~\bibnamefont
  {{Gooth}}}, \bibinfo {author} {\bibfnamefont {F.}~\bibnamefont {{Menges}}},
  \bibinfo {author} {\bibfnamefont {N.}~\bibnamefont {{Kumar}}}, \bibinfo
  {author} {\bibfnamefont {V.}~\bibnamefont {{S{\"u}{\ss}}}}, \bibinfo {author}
  {\bibfnamefont {C.}~\bibnamefont {{Shekhar}}}, \bibinfo {author}
  {\bibfnamefont {Y.}~\bibnamefont {{Sun}}}, \bibinfo {author} {\bibfnamefont
  {U.}~\bibnamefont {{Drechsler}}}, \bibinfo {author} {\bibfnamefont
  {R.}~\bibnamefont {{Zierold}}}, \bibinfo {author} {\bibfnamefont
  {C.}~\bibnamefont {{Felser}}}, \ and\ \bibinfo {author} {\bibfnamefont
  {B.}~\bibnamefont {{Gotsmann}}},\ }\href {\doibase
  10.1038/s41467-018-06688-y} {\bibfield  {journal} {\bibinfo  {journal} {Nat.
  Comm.}\ }\textbf {\bibinfo {volume} {9}},\ \bibinfo {pages} {4093} (\bibinfo
  {year} {2018})}\BibitemShut {NoStop}%
\bibitem [{\citenamefont {{Berdyugin}}\ \emph {et~al.}(2019)\citenamefont
  {{Berdyugin}}, \citenamefont {{Xu}}, \citenamefont {{Pellegrino}},
  \citenamefont {{Krishna Kumar}}, \citenamefont {{Principi}}, \citenamefont
  {{Torre}}, \citenamefont {{Ben Shalom}}, \citenamefont {{Taniguchi}},
  \citenamefont {{Watanabe}}, \citenamefont {{Grigorieva}}, \citenamefont
  {{Polini}}, \citenamefont {{Geim}},\ and\ \citenamefont
  {{Bandurin}}}]{Berdyugin2019}%
  \BibitemOpen
  \bibfield  {author} {\bibinfo {author} {\bibfnamefont {A.~I.}\ \bibnamefont
  {{Berdyugin}}}, \bibinfo {author} {\bibfnamefont {S.~G.}\ \bibnamefont
  {{Xu}}}, \bibinfo {author} {\bibfnamefont {F.~M.~D.}\ \bibnamefont
  {{Pellegrino}}}, \bibinfo {author} {\bibfnamefont {R.}~\bibnamefont {{Krishna
  Kumar}}}, \bibinfo {author} {\bibfnamefont {A.}~\bibnamefont {{Principi}}},
  \bibinfo {author} {\bibfnamefont {I.}~\bibnamefont {{Torre}}}, \bibinfo
  {author} {\bibfnamefont {M.}~\bibnamefont {{Ben Shalom}}}, \bibinfo {author}
  {\bibfnamefont {T.}~\bibnamefont {{Taniguchi}}}, \bibinfo {author}
  {\bibfnamefont {K.}~\bibnamefont {{Watanabe}}}, \bibinfo {author}
  {\bibfnamefont {I.~V.}\ \bibnamefont {{Grigorieva}}}, \bibinfo {author}
  {\bibfnamefont {M.}~\bibnamefont {{Polini}}}, \bibinfo {author}
  {\bibfnamefont {A.~K.}\ \bibnamefont {{Geim}}}, \ and\ \bibinfo {author}
  {\bibfnamefont {D.~A.}\ \bibnamefont {{Bandurin}}},\ }\href {\doibase
  10.1126/science.aau0685} {\bibfield  {journal} {\bibinfo  {journal}
  {Science}\ }\textbf {\bibinfo {volume} {364}},\ \bibinfo {pages} {162}
  (\bibinfo {year} {2019})}\BibitemShut {NoStop}%
\bibitem [{\citenamefont {{Andreev}}\ \emph {et~al.}(2011)\citenamefont
  {{Andreev}}, \citenamefont {{Kivelson}},\ and\ \citenamefont
  {{Spivak}}}]{Andreev2011}%
  \BibitemOpen
  \bibfield  {author} {\bibinfo {author} {\bibfnamefont {A.~V.}\ \bibnamefont
  {{Andreev}}}, \bibinfo {author} {\bibfnamefont {S.~A.}\ \bibnamefont
  {{Kivelson}}}, \ and\ \bibinfo {author} {\bibfnamefont {B.}~\bibnamefont
  {{Spivak}}},\ }\href {\doibase 10.1103/PhysRevLett.106.256804} {\bibfield
  {journal} {\bibinfo  {journal} {Phys. Rev. Lett.}\ }\textbf {\bibinfo
  {volume} {106}},\ \bibinfo {pages} {256804} (\bibinfo {year}
  {2011})}\BibitemShut {NoStop}%
\bibitem [{\citenamefont {{Tomadin}}\ \emph {et~al.}(2014)\citenamefont
  {{Tomadin}}, \citenamefont {{Vignale}},\ and\ \citenamefont
  {{Polini}}}]{Tomadin2014}%
  \BibitemOpen
  \bibfield  {author} {\bibinfo {author} {\bibfnamefont {A.}~\bibnamefont
  {{Tomadin}}}, \bibinfo {author} {\bibfnamefont {G.}~\bibnamefont
  {{Vignale}}}, \ and\ \bibinfo {author} {\bibfnamefont {M.}~\bibnamefont
  {{Polini}}},\ }\href {\doibase 10.1103/PhysRevLett.113.235901} {\bibfield
  {journal} {\bibinfo  {journal} {Phys. Rev. Lett.}\ }\textbf {\bibinfo
  {volume} {113}},\ \bibinfo {pages} {235901} (\bibinfo {year}
  {2014})}\BibitemShut {NoStop}%
\bibitem [{\citenamefont {{Torre}}\ \emph {et~al.}(2015)\citenamefont
  {{Torre}}, \citenamefont {{Tomadin}}, \citenamefont {{Geim}},\ and\
  \citenamefont {{Polini}}}]{Torre2015}%
  \BibitemOpen
  \bibfield  {author} {\bibinfo {author} {\bibfnamefont {I.}~\bibnamefont
  {{Torre}}}, \bibinfo {author} {\bibfnamefont {A.}~\bibnamefont {{Tomadin}}},
  \bibinfo {author} {\bibfnamefont {A.~K.}\ \bibnamefont {{Geim}}}, \ and\
  \bibinfo {author} {\bibfnamefont {M.}~\bibnamefont {{Polini}}},\ }\href
  {\doibase 10.1103/PhysRevB.92.165433} {\bibfield  {journal} {\bibinfo
  {journal} {Phys. Rev. B}\ }\textbf {\bibinfo {volume} {92}},\ \bibinfo
  {pages} {165433} (\bibinfo {year} {2015})}\BibitemShut {NoStop}%
\bibitem [{\citenamefont {{Levchenko}}\ \emph {et~al.}(2017)\citenamefont
  {{Levchenko}}, \citenamefont {{Xie}},\ and\ \citenamefont
  {{Andreev}}}]{Levchenko2017}%
  \BibitemOpen
  \bibfield  {author} {\bibinfo {author} {\bibfnamefont {A.}~\bibnamefont
  {{Levchenko}}}, \bibinfo {author} {\bibfnamefont {H.-Y.}\ \bibnamefont
  {{Xie}}}, \ and\ \bibinfo {author} {\bibfnamefont {A.~V.}\ \bibnamefont
  {{Andreev}}},\ }\href {\doibase 10.1103/PhysRevB.95.121301} {\bibfield
  {journal} {\bibinfo  {journal} {Phys. Rev. B}\ }\textbf {\bibinfo {volume}
  {95}},\ \bibinfo {pages} {121301(R)} (\bibinfo {year} {2017})}\BibitemShut
  {NoStop}%
\bibitem [{\citenamefont {{Svintsov}}(2018)}]{Svintsov2018}%
  \BibitemOpen
  \bibfield  {author} {\bibinfo {author} {\bibfnamefont {D.}~\bibnamefont
  {{Svintsov}}},\ }\href {\doibase 10.1103/PhysRevB.97.121405} {\bibfield
  {journal} {\bibinfo  {journal} {Phys. Rev. B}\ }\textbf {\bibinfo {volume}
  {97}},\ \bibinfo {pages} {121405(R)} (\bibinfo {year} {2018})}\BibitemShut
  {NoStop}%
\bibitem [{\citenamefont {{Ho}}\ \emph {et~al.}(2018)\citenamefont {{Ho}},
  \citenamefont {{Yudhistira}}, \citenamefont {{Chakraborty}},\ and\
  \citenamefont {{Adam}}}]{Ho2018}%
  \BibitemOpen
  \bibfield  {author} {\bibinfo {author} {\bibfnamefont {D.~Y.~H.}\
  \bibnamefont {{Ho}}}, \bibinfo {author} {\bibfnamefont {I.}~\bibnamefont
  {{Yudhistira}}}, \bibinfo {author} {\bibfnamefont {N.}~\bibnamefont
  {{Chakraborty}}}, \ and\ \bibinfo {author} {\bibfnamefont {S.}~\bibnamefont
  {{Adam}}},\ }\href {\doibase 10.1103/PhysRevB.97.121404} {\bibfield
  {journal} {\bibinfo  {journal} {Phys. Rev. B}\ }\textbf {\bibinfo {volume}
  {97}},\ \bibinfo {pages} {121404(R)} (\bibinfo {year} {2018})}\BibitemShut
  {NoStop}%
\bibitem [{\citenamefont {{Lucas}}\ and\ \citenamefont
  {{Sur{\'o}wka}}(2014)}]{Lucas2014}%
  \BibitemOpen
  \bibfield  {author} {\bibinfo {author} {\bibfnamefont {A.}~\bibnamefont
  {{Lucas}}}\ and\ \bibinfo {author} {\bibfnamefont {P.}~\bibnamefont
  {{Sur{\'o}wka}}},\ }\href {\doibase 10.1103/PhysRevE.90.063005} {\bibfield
  {journal} {\bibinfo  {journal} {Phys. Rev. E}\ }\textbf {\bibinfo {volume}
  {90}},\ \bibinfo {pages} {063005} (\bibinfo {year} {2014})}\BibitemShut
  {NoStop}%
\bibitem [{\citenamefont {{Sherafati}}\ \emph {et~al.}(2016)\citenamefont
  {{Sherafati}}, \citenamefont {{Principi}},\ and\ \citenamefont
  {{Vignale}}}]{Sherafati2016}%
  \BibitemOpen
  \bibfield  {author} {\bibinfo {author} {\bibfnamefont {M.}~\bibnamefont
  {{Sherafati}}}, \bibinfo {author} {\bibfnamefont {A.}~\bibnamefont
  {{Principi}}}, \ and\ \bibinfo {author} {\bibfnamefont {G.}~\bibnamefont
  {{Vignale}}},\ }\href {\doibase 10.1103/PhysRevB.94.125427} {\bibfield
  {journal} {\bibinfo  {journal} {Phys. Rev. B}\ }\textbf {\bibinfo {volume}
  {94}},\ \bibinfo {pages} {125427} (\bibinfo {year} {2016})}\BibitemShut
  {NoStop}%
\bibitem [{\citenamefont {{Ganeshan}}\ and\ \citenamefont
  {{Abanov}}(2017)}]{Ganeshan2017}%
  \BibitemOpen
  \bibfield  {author} {\bibinfo {author} {\bibfnamefont {S.}~\bibnamefont
  {{Ganeshan}}}\ and\ \bibinfo {author} {\bibfnamefont {A.~G.}\ \bibnamefont
  {{Abanov}}},\ }\href {\doibase 10.1103/PhysRevFluids.2.094101} {\bibfield
  {journal} {\bibinfo  {journal} {Phys. Rev. Fluids}\ }\textbf {\bibinfo
  {volume} {2}},\ \bibinfo {pages} {094101} (\bibinfo {year}
  {2017})}\BibitemShut {NoStop}%
\bibitem [{\citenamefont {Guerrero-Becerra}\ \emph {et~al.}(2019)\citenamefont
  {Guerrero-Becerra}, \citenamefont {Pellegrino},\ and\ \citenamefont
  {Polini}}]{Guerrero-Becerra2018}%
  \BibitemOpen
  \bibfield  {author} {\bibinfo {author} {\bibfnamefont {K.~A.}\ \bibnamefont
  {Guerrero-Becerra}}, \bibinfo {author} {\bibfnamefont {F.~M.~D.}\
  \bibnamefont {Pellegrino}}, \ and\ \bibinfo {author} {\bibfnamefont
  {M.}~\bibnamefont {Polini}},\ }\href {\doibase 10.1103/PhysRevB.99.041407}
  {\bibfield  {journal} {\bibinfo  {journal} {Phys. Rev. B}\ }\textbf {\bibinfo
  {volume} {99}},\ \bibinfo {pages} {041407(R)} (\bibinfo {year}
  {2019})}\BibitemShut {NoStop}%
\bibitem [{\citenamefont {{Gusev}}\ \emph {et~al.}(2018)\citenamefont
  {{Gusev}}, \citenamefont {{Levin}}, \citenamefont {{Levinson}},\ and\
  \citenamefont {{Bakarov}}}]{Gusev2018}%
  \BibitemOpen
  \bibfield  {author} {\bibinfo {author} {\bibfnamefont {G.~M.}\ \bibnamefont
  {{Gusev}}}, \bibinfo {author} {\bibfnamefont {A.~D.}\ \bibnamefont
  {{Levin}}}, \bibinfo {author} {\bibfnamefont {E.~V.}\ \bibnamefont
  {{Levinson}}}, \ and\ \bibinfo {author} {\bibfnamefont {A.~K.}\ \bibnamefont
  {{Bakarov}}},\ }\href {\doibase 10.1103/PhysRevB.98.161303} {\bibfield
  {journal} {\bibinfo  {journal} {Phys. Rev. B}\ }\textbf {\bibinfo {volume}
  {98}},\ \bibinfo {pages} {161303(R)} (\bibinfo {year} {2018})}\BibitemShut
  {NoStop}%
\bibitem [{\citenamefont {{Lapa}}\ and\ \citenamefont
  {{Hughes}}(2014)}]{Lapa2014}%
  \BibitemOpen
  \bibfield  {author} {\bibinfo {author} {\bibfnamefont {M.~F.}\ \bibnamefont
  {{Lapa}}}\ and\ \bibinfo {author} {\bibfnamefont {T.~L.}\ \bibnamefont
  {{Hughes}}},\ }\href {\doibase 10.1103/PhysRevE.89.043019} {\bibfield
  {journal} {\bibinfo  {journal} {Phys. Rev. E}\ }\textbf {\bibinfo {volume}
  {89}},\ \bibinfo {pages} {043019} (\bibinfo {year} {2014})}\BibitemShut
  {NoStop}%
\bibitem [{\citenamefont {{Tuegel}}\ and\ \citenamefont
  {{Hughes}}(2017)}]{Tuegel2017}%
  \BibitemOpen
  \bibfield  {author} {\bibinfo {author} {\bibfnamefont {T.~I.}\ \bibnamefont
  {{Tuegel}}}\ and\ \bibinfo {author} {\bibfnamefont {T.~L.}\ \bibnamefont
  {{Hughes}}},\ }\href {\doibase 10.1103/PhysRevB.96.174524} {\bibfield
  {journal} {\bibinfo  {journal} {Phys. Rev. B}\ }\textbf {\bibinfo {volume}
  {96}},\ \bibinfo {pages} {174524} (\bibinfo {year} {2017})}\BibitemShut
  {NoStop}%
\bibitem [{\citenamefont {{Banerjee}}\ \emph {et~al.}(2017)\citenamefont
  {{Banerjee}}, \citenamefont {{Souslov}}, \citenamefont {{Abanov}},\ and\
  \citenamefont {{Vitelli}}}]{Banerjee2017}%
  \BibitemOpen
  \bibfield  {author} {\bibinfo {author} {\bibfnamefont {D.}~\bibnamefont
  {{Banerjee}}}, \bibinfo {author} {\bibfnamefont {A.}~\bibnamefont
  {{Souslov}}}, \bibinfo {author} {\bibfnamefont {A.~G.}\ \bibnamefont
  {{Abanov}}}, \ and\ \bibinfo {author} {\bibfnamefont {V.}~\bibnamefont
  {{Vitelli}}},\ }\href {\doibase 10.1038/s41467-017-01378-7} {\bibfield
  {journal} {\bibinfo  {journal} {Nat. Comm.}\ }\textbf {\bibinfo {volume}
  {8}},\ \bibinfo {pages} {1573} (\bibinfo {year} {2017})}\BibitemShut
  {NoStop}%
\bibitem [{\citenamefont {{Soni}}\ \emph {et~al.}(2018)\citenamefont {{Soni}},
  \citenamefont {{Bililign}}, \citenamefont {{Magkiriadou}}, \citenamefont
  {{Sacanna}}, \citenamefont {{Bartolo}}, \citenamefont {{Shelley}},\ and\
  \citenamefont {{Irvine}}}]{Soni2018}%
  \BibitemOpen
  \bibfield  {author} {\bibinfo {author} {\bibfnamefont {V.}~\bibnamefont
  {{Soni}}}, \bibinfo {author} {\bibfnamefont {E.}~\bibnamefont {{Bililign}}},
  \bibinfo {author} {\bibfnamefont {S.}~\bibnamefont {{Magkiriadou}}}, \bibinfo
  {author} {\bibfnamefont {S.}~\bibnamefont {{Sacanna}}}, \bibinfo {author}
  {\bibfnamefont {D.}~\bibnamefont {{Bartolo}}}, \bibinfo {author}
  {\bibfnamefont {M.~J.}\ \bibnamefont {{Shelley}}}, \ and\ \bibinfo {author}
  {\bibfnamefont {W.~T.~M.}\ \bibnamefont {{Irvine}}},\ }\href@noop {}
  {\bibfield  {journal} {\bibinfo  {journal} {arXiv}\ ,\ \bibinfo {pages}
  {1812.09990}} (\bibinfo {year} {2018})}\BibitemShut {NoStop}%
\bibitem [{\citenamefont {Souslov}\ \emph {et~al.}(2019)\citenamefont
  {Souslov}, \citenamefont {Dasbiswas}, \citenamefont {Fruchart}, \citenamefont
  {Vaikuntanathan},\ and\ \citenamefont {Vitelli}}]{Souslov2019}%
  \BibitemOpen
  \bibfield  {author} {\bibinfo {author} {\bibfnamefont {A.}~\bibnamefont
  {Souslov}}, \bibinfo {author} {\bibfnamefont {K.}~\bibnamefont {Dasbiswas}},
  \bibinfo {author} {\bibfnamefont {M.}~\bibnamefont {Fruchart}}, \bibinfo
  {author} {\bibfnamefont {S.}~\bibnamefont {Vaikuntanathan}}, \ and\ \bibinfo
  {author} {\bibfnamefont {V.}~\bibnamefont {Vitelli}},\ }\href {\doibase
  10.1103/PhysRevLett.122.128001} {\bibfield  {journal} {\bibinfo  {journal}
  {Phys. Rev. Lett.}\ }\textbf {\bibinfo {volume} {122}},\ \bibinfo {pages}
  {128001} (\bibinfo {year} {2019})}\BibitemShut {NoStop}%
\bibitem [{\citenamefont {Ella}\ \emph {et~al.}(2019)\citenamefont {Ella},
  \citenamefont {Rozen}, \citenamefont {Birkbeck}, \citenamefont {Ben-Shalom},
  \citenamefont {Perello}, \citenamefont {Zultak}, \citenamefont {Taniguchi},
  \citenamefont {Watanabe}, \citenamefont {Geim}, \citenamefont {Ilani},\ and\
  \citenamefont {Sulpizio}}]{Ella2018}%
  \BibitemOpen
  \bibfield  {author} {\bibinfo {author} {\bibfnamefont {L.}~\bibnamefont
  {Ella}}, \bibinfo {author} {\bibfnamefont {A.}~\bibnamefont {Rozen}},
  \bibinfo {author} {\bibfnamefont {J.}~\bibnamefont {Birkbeck}}, \bibinfo
  {author} {\bibfnamefont {M.}~\bibnamefont {Ben-Shalom}}, \bibinfo {author}
  {\bibfnamefont {D.}~\bibnamefont {Perello}}, \bibinfo {author} {\bibfnamefont
  {J.}~\bibnamefont {Zultak}}, \bibinfo {author} {\bibfnamefont
  {T.}~\bibnamefont {Taniguchi}}, \bibinfo {author} {\bibfnamefont
  {K.}~\bibnamefont {Watanabe}}, \bibinfo {author} {\bibfnamefont {A.~K.}\
  \bibnamefont {Geim}}, \bibinfo {author} {\bibfnamefont {S.}~\bibnamefont
  {Ilani}}, \ and\ \bibinfo {author} {\bibfnamefont {J.~A.}\ \bibnamefont
  {Sulpizio}},\ }\href {\doibase 10.1038/s41565-019-0398-x} {\bibfield
  {journal} {\bibinfo  {journal} {Nat. Nanotechnol.}\ }\textbf {\bibinfo
  {volume} {14}},\ \bibinfo {pages} {480} (\bibinfo {year} {2019})}\BibitemShut
  {NoStop}%
\bibitem [{\citenamefont {{Sulpizio}}\ \emph {et~al.}(2019)\citenamefont
  {{Sulpizio}}, \citenamefont {{Ella}}, \citenamefont {{Rozen}}, \citenamefont
  {{Birkbeck}}, \citenamefont {{Perello}}, \citenamefont {{Dutta}},
  \citenamefont {{Ben-Shalom}}, \citenamefont {{Taniguchi}}, \citenamefont
  {{Watanabe}}, \citenamefont {{Holder}}, \citenamefont {{Queiroz}},
  \citenamefont {{Stern}}, \citenamefont {{Scaffidi}}, \citenamefont {{Geim}},\
  and\ \citenamefont {{Ilani}}}]{Sulpizio2019}%
  \BibitemOpen
  \bibfield  {author} {\bibinfo {author} {\bibfnamefont {J.~A.}\ \bibnamefont
  {{Sulpizio}}}, \bibinfo {author} {\bibfnamefont {L.}~\bibnamefont {{Ella}}},
  \bibinfo {author} {\bibfnamefont {A.}~\bibnamefont {{Rozen}}}, \bibinfo
  {author} {\bibfnamefont {J.}~\bibnamefont {{Birkbeck}}}, \bibinfo {author}
  {\bibfnamefont {D.~J.}\ \bibnamefont {{Perello}}}, \bibinfo {author}
  {\bibfnamefont {D.}~\bibnamefont {{Dutta}}}, \bibinfo {author} {\bibfnamefont
  {M.}~\bibnamefont {{Ben-Shalom}}}, \bibinfo {author} {\bibfnamefont
  {T.}~\bibnamefont {{Taniguchi}}}, \bibinfo {author} {\bibfnamefont
  {K.}~\bibnamefont {{Watanabe}}}, \bibinfo {author} {\bibfnamefont
  {T.}~\bibnamefont {{Holder}}}, \bibinfo {author} {\bibfnamefont
  {R.}~\bibnamefont {{Queiroz}}}, \bibinfo {author} {\bibfnamefont
  {A.}~\bibnamefont {{Stern}}}, \bibinfo {author} {\bibfnamefont
  {T.}~\bibnamefont {{Scaffidi}}}, \bibinfo {author} {\bibfnamefont {A.~K.}\
  \bibnamefont {{Geim}}}, \ and\ \bibinfo {author} {\bibfnamefont
  {S.}~\bibnamefont {{Ilani}}},\ }\href@noop {} {\bibfield  {journal} {\bibinfo
   {journal} {arXiv}\ ,\ \bibinfo {eid} {arXiv:1905.11662}} (\bibinfo {year}
  {2019})}\BibitemShut {NoStop}%
\bibitem [{\citenamefont {{Alekseev}}(2016)}]{Alekseev2016}%
  \BibitemOpen
  \bibfield  {author} {\bibinfo {author} {\bibfnamefont {P.~S.}\ \bibnamefont
  {{Alekseev}}},\ }\href {\doibase 10.1103/PhysRevLett.117.166601} {\bibfield
  {journal} {\bibinfo  {journal} {Phys. Rev. Lett.}\ }\textbf {\bibinfo
  {volume} {117}},\ \bibinfo {pages} {166601} (\bibinfo {year}
  {2016})}\BibitemShut {NoStop}%
\bibitem [{\citenamefont {{Falkovich}}\ and\ \citenamefont
  {{Levitov}}(2017)}]{Falkovich2017}%
  \BibitemOpen
  \bibfield  {author} {\bibinfo {author} {\bibfnamefont {G.}~\bibnamefont
  {{Falkovich}}}\ and\ \bibinfo {author} {\bibfnamefont {L.}~\bibnamefont
  {{Levitov}}},\ }\href {\doibase 10.1103/PhysRevLett.119.066601} {\bibfield
  {journal} {\bibinfo  {journal} {Phys. Rev. Lett.}\ }\textbf {\bibinfo
  {volume} {119}},\ \bibinfo {pages} {066601} (\bibinfo {year}
  {2017})}\BibitemShut {NoStop}%
\bibitem [{\citenamefont {{Delacr{\'e}taz}}\ and\ \citenamefont
  {{Gromov}}(2017)}]{Delacretaz2017}%
  \BibitemOpen
  \bibfield  {author} {\bibinfo {author} {\bibfnamefont {L.~V.}\ \bibnamefont
  {{Delacr{\'e}taz}}}\ and\ \bibinfo {author} {\bibfnamefont {A.}~\bibnamefont
  {{Gromov}}},\ }\href {\doibase 10.1103/PhysRevLett.119.226602} {\bibfield
  {journal} {\bibinfo  {journal} {Phys. Rev. Lett.}\ }\textbf {\bibinfo
  {volume} {119}},\ \bibinfo {pages} {226602} (\bibinfo {year}
  {2017})}\BibitemShut {NoStop}%
\bibitem [{Note1()}]{Note1}%
  \BibitemOpen
  \bibinfo {note} {The angular momentum components of the distribution function
  are defined by $h^{e}_l(y)=\pi ^{-1}\DOTSI \intop \ilimits@ h(\theta
  ,y)\protect \qopname \relax o{cos}(l\theta ) d\theta $ and $h^{o}_l(y)=\pi
  ^{-1}\DOTSI \intop \ilimits@ h(\theta ,y)\protect \qopname \relax
  o{sin}(l\theta ) d\theta $}\BibitemShut {NoStop}%
\bibitem [{\citenamefont {Lifshitz}\ and\ \citenamefont
  {Pitaevskii}(1980)}]{Landaubook9}%
  \BibitemOpen
  \bibfield  {author} {\bibinfo {author} {\bibfnamefont {E.}~\bibnamefont
  {Lifshitz}}\ and\ \bibinfo {author} {\bibfnamefont {L.}~\bibnamefont
  {Pitaevskii}},\ }\href@noop {} {\emph {\bibinfo {title} {Statistical Physics,
  Part 2}}},\ \bibinfo {series} {Course of Theoretical Physics}\ No.~\bibinfo
  {number} {9}\ (\bibinfo  {publisher} {Pergamon},\ \bibinfo {address} {New
  York},\ \bibinfo {year} {1980})\BibitemShut {NoStop}%
\bibitem [{Note2()}]{Note2}%
  \BibitemOpen
  \bibinfo {note} {The most famous example for a Fermi liquid with strong
  interactions is undoubtedly Graphene at finite density, where both theory and
  experiment estimate $\ell _{ee}\sim 1\protect \mathrm {\mu m}\ll w$ at
  temperature $T=75\protect \mathrm {K}$~\cite
  {Schuett2011,Svintsov2018,Ho2018,Bandurin2016,KrishnaKumar2017}.}\BibitemShut
  {Stop}%
\bibitem [{Note3()}]{Note3}%
  \BibitemOpen
  \bibinfo {note} {The stress tensor is closely related to $h_2^{e/o}$.
  Disregarding charging effects, stress tensor and second moments can be used
  interchangeably.}\BibitemShut {Stop}%
\bibitem [{Note4()}]{Note4}%
  \BibitemOpen
  \bibinfo {note} {We point out that at zero magnetic field this length
  approaches the Gurzhi length, $\ell _c\rightarrow \protect \sqrt {\ell \ell
  _0}$.}\BibitemShut {Stop}%
\bibitem [{\citenamefont {Kiselev}\ and\ \citenamefont
  {Schmalian}(2019)}]{Kiselev2018}%
  \BibitemOpen
  \bibfield  {author} {\bibinfo {author} {\bibfnamefont {E.~I.}\ \bibnamefont
  {Kiselev}}\ and\ \bibinfo {author} {\bibfnamefont {J.}~\bibnamefont
  {Schmalian}},\ }\href {\doibase 10.1103/PhysRevB.99.035430} {\bibfield
  {journal} {\bibinfo  {journal} {Phys. Rev. B}\ }\textbf {\bibinfo {volume}
  {99}},\ \bibinfo {pages} {035430} (\bibinfo {year} {2019})}\BibitemShut
  {NoStop}%
\bibitem [{\citenamefont {{Scaffidi}}\ \emph {et~al.}(2017)\citenamefont
  {{Scaffidi}}, \citenamefont {{Nandi}}, \citenamefont {{Schmidt}},
  \citenamefont {{Mackenzie}},\ and\ \citenamefont {{Moore}}}]{Scaffidi2017}%
  \BibitemOpen
  \bibfield  {author} {\bibinfo {author} {\bibfnamefont {T.}~\bibnamefont
  {{Scaffidi}}}, \bibinfo {author} {\bibfnamefont {N.}~\bibnamefont {{Nandi}}},
  \bibinfo {author} {\bibfnamefont {B.}~\bibnamefont {{Schmidt}}}, \bibinfo
  {author} {\bibfnamefont {A.~P.}\ \bibnamefont {{Mackenzie}}}, \ and\ \bibinfo
  {author} {\bibfnamefont {J.~E.}\ \bibnamefont {{Moore}}},\ }\href {\doibase
  10.1103/PhysRevLett.118.226601} {\bibfield  {journal} {\bibinfo  {journal}
  {Phys. Rev. Lett.}\ }\textbf {\bibinfo {volume} {118}},\ \bibinfo {pages}
  {226601} (\bibinfo {year} {2017})}\BibitemShut {NoStop}%
\bibitem [{\citenamefont {{Alekseev}}\ and\ \citenamefont
  {{Semina}}(2018)}]{Alekseev2018}%
  \BibitemOpen
  \bibfield  {author} {\bibinfo {author} {\bibfnamefont {P.~S.}\ \bibnamefont
  {{Alekseev}}}\ and\ \bibinfo {author} {\bibfnamefont {M.~A.}\ \bibnamefont
  {{Semina}}},\ }\href {\doibase 10.1103/PhysRevB.98.165412} {\bibfield
  {journal} {\bibinfo  {journal} {Phys. Rev. B}\ }\textbf {\bibinfo {volume}
  {98}},\ \bibinfo {pages} {165412} (\bibinfo {year} {2018})}\BibitemShut
  {NoStop}%
\bibitem [{\citenamefont {{Beenakker}}\ and\ \citenamefont {{van
  Houten}}(1989)}]{Beenakker1989}%
  \BibitemOpen
  \bibfield  {author} {\bibinfo {author} {\bibfnamefont {C.~W.~J.}\
  \bibnamefont {{Beenakker}}}\ and\ \bibinfo {author} {\bibfnamefont
  {H.}~\bibnamefont {{van Houten}}},\ }\href {\doibase
  10.1103/PhysRevLett.63.1857} {\bibfield  {journal} {\bibinfo  {journal}
  {Phys. Rev. Lett.}\ }\textbf {\bibinfo {volume} {63}},\ \bibinfo {pages}
  {1857} (\bibinfo {year} {1989})}\BibitemShut {NoStop}%
\bibitem [{\citenamefont {{Soffer}}(1967)}]{Soffer1967}%
  \BibitemOpen
  \bibfield  {author} {\bibinfo {author} {\bibfnamefont {S.~B.}\ \bibnamefont
  {{Soffer}}},\ }\href {\doibase 10.1063/1.1709746} {\bibfield  {journal}
  {\bibinfo  {journal} {J. Appl. Phys.}\ }\textbf {\bibinfo {volume} {38}},\
  \bibinfo {pages} {1710} (\bibinfo {year} {1967})}\BibitemShut {NoStop}%
\bibitem [{\citenamefont {{Lucas}}(2017)}]{Lucas2017}%
  \BibitemOpen
  \bibfield  {author} {\bibinfo {author} {\bibfnamefont {A.}~\bibnamefont
  {{Lucas}}},\ }\href@noop {} {\bibfield  {journal} {\bibinfo  {journal}
  {arXiv}\ ,\ \bibinfo {pages} {1710.01005}} (\bibinfo {year}
  {2017})}\BibitemShut {NoStop}%
\bibitem [{\citenamefont {{Pandey}}(1996)}]{Pandey1996}%
  \BibitemOpen
  \bibfield  {author} {\bibinfo {author} {\bibfnamefont {J.~N.}\ \bibnamefont
  {{Pandey}}},\ }\href@noop {} {\emph {\bibinfo {title} {The Hilbert transform
  of Schwartz distributions and applications}}}\ (\bibinfo  {publisher}
  {Wiley-Interscience},\ \bibinfo {year} {1996})\BibitemShut {NoStop}%
\bibitem [{\citenamefont {{Sch{\"u}tt}}\ \emph {et~al.}(2011)\citenamefont
  {{Sch{\"u}tt}}, \citenamefont {{Ostrovsky}}, \citenamefont {{Gornyi}},\ and\
  \citenamefont {{Mirlin}}}]{Schuett2011}%
  \BibitemOpen
  \bibfield  {author} {\bibinfo {author} {\bibfnamefont {M.}~\bibnamefont
  {{Sch{\"u}tt}}}, \bibinfo {author} {\bibfnamefont {P.~M.}\ \bibnamefont
  {{Ostrovsky}}}, \bibinfo {author} {\bibfnamefont {I.~V.}\ \bibnamefont
  {{Gornyi}}}, \ and\ \bibinfo {author} {\bibfnamefont {A.~D.}\ \bibnamefont
  {{Mirlin}}},\ }\href {\doibase 10.1103/PhysRevB.83.155441} {\bibfield
  {journal} {\bibinfo  {journal} {Phys. Rev. B}\ }\textbf {\bibinfo {volume}
  {83}},\ \bibinfo {eid} {155441} (\bibinfo {year} {2011})}\BibitemShut
  {NoStop}%
\end{thebibliography}
\end{document}